\documentclass[jcp,twocolumn,showpacs,superscriptaddress]{revtex4-1}

\usepackage[utf8]{inputenc}
\usepackage[T1]{fontenc}

\usepackage{multirow}	
\usepackage{enumerate}	
\usepackage{Tabbing}	
\usepackage{relsize}	
\usepackage{textcomp}	
\usepackage{soul}	

\usepackage{amsmath}
\usepackage{amssymb}
\usepackage{bm}		

\usepackage{hyperref}
\usepackage{url}

\usepackage{dcolumn}	
\usepackage{longtable}	
\usepackage{booktabs}	

\usepackage[pdftex]{graphicx}
\usepackage{epstopdf}	
\usepackage{rotating}	
\usepackage{float}	

\usepackage{color}

\newcommand{\ket}[1]{\ensuremath{\left|#1\right\rangle}}
\newcommand{\average}[1]{\ensuremath{\left\langle#1\right\rangle}}
\newcommand{\bracket}[2]{\ensuremath{\left\langle#1 \vphantom{#2}\right| \left. #2 \vphantom{#1}\right\rangle}}
\newcommand{\matrixel}[3]{\ensuremath{\left\langle #1 \vphantom{#2#3} \right| #2 \left| #3 \vphantom{#1#2} \right\rangle}}

\newcommand{\AN}[2]{
\ensuremath{\hat{a}^{} _{{#1}
\ifnum#2=1  \uparrow
\else
\ifnum#2=-1  \downarrow
\else
\ifnum#2=2  \sigma
\else
\ifnum#2=-2  \bar{\sigma}
\fi
\fi
\fi
\fi
}}
}
\newcommand{\CR}[2]{
\ensuremath{\hat{a}^\dagger _{{#1}
\ifnum#2=1  \uparrow
\else
\ifnum#2=-1  \downarrow
\else
\ifnum#2=2  \sigma
\else
\ifnum#2=-2  \bar{\sigma}
\fi
\fi
\fi
\fi
}}
}
\newcommand{\NUM}[2]{
\ensuremath{\hat{n}^{} _{{#1}
\ifnum#2=1  \uparrow
\else
\ifnum#2=-1  \downarrow
\else
\ifnum#2=2  \sigma
\else
\ifnum#2=-2  \bar{\sigma}
\fi
\fi
\fi
\fi
}}
}

\newcommand{\mc}[3]{\multicolumn{#1}{#2}{#3}}

\newcommand{\Ang}{
\text{\AA{}}
}

\renewcommand{\vec}[1]{\ensuremath{\mathbf{#1}}}

\newcommand{\figuresize}{0.99\linewidth}

\newcommand{\wyrozn}[1]{#1}
\newcommand{\Slater}{\ensuremath{'}}

\begin{document}

\title{$H_2$ and $\left( H_2 \right) _2$ molecules with an \textit{ab initio} optimization of wave functions in correlated state: Electron--proton couplings and intermolecular microscopic parameters}

\author{Andrzej P. K\k{a}dzielawa}
\email{kadzielawa@th.if.uj.edu.pl}
\affiliation{                                                                 
Instytut Fizyki im. Mariana Smoluchowskiego, Uniwersytet Jagielloński, ul. Reymonta 4, PL-30059 Krak\'{o}w, Poland
}
\author{Agata Bielas}
\affiliation{
Instytut Fizyki, Uniwersytet Śląski, ul. Uniwersytecka 4, PL-40007 Katowice, Poland
}
\author{Marcello Acquarone}
\affiliation{
Dipartimento di Fisica e Scienze della Terra dell’Universit\`{a} di Parma, I-43100 Parma, Italy
}
\author{Andrzej Biborski}
\affiliation{
Akademickie Centrum Materiałów i Nanotechnologii,
AGH Akademia Górniczo-Hutnicza,
Al. Mickiewicza 30,
PL-30-059 Kraków
}
\author{Maciej M. Ma\'{s}ka}
\affiliation{
Instytut Fizyki, Uniwersytet Śląski, ul. Uniwersytecka 4, PL-40007 Katowice, Poland
}
\author{J\'{o}zef Spa\l{}ek}
\email{ufspalek@if.uj.edu.pl}
\affiliation{                                                                 
Instytut Fizyki im. Mariana Smoluchowskiego, Uniwersytet Jagielloński, ul. Reymonta 4, PL-30059 Krak\'{o}w, Poland
}
\affiliation{
Akademickie Centrum Materiałów i Nanotechnologii,
AGH Akademia Górniczo-Hutnicza,
Al. Mickiewicza 30,
PL-30-059 Kraków
}

\date{July 3, 2014}

\begin{abstract}
The hydrogen molecules $H_2$ and $\left( H_2 \right) _2$ are analyzed with electronic correlations taken into account between the $1s$ electrons in an exact manner.
The optimal single-particle
Slater orbitals are evaluated in the correlated state of $H_2$ by combining their variational determination with the diagonalization of~the~full Hamiltonian in
the~second-quantization language. All electron--ion coupling constants are determined explicitly and their relative importance is discussed. 
Sizable zero-point motion amplitude and the corresponding
energy are then evaluated by taking into account the anharmonic contributions up to the ninth order in the relative displacement of the ions from their static equilibrium
value. The applicability of the model to the solid molecular hydrogen is briefly analyzed by calculating intermolecular microscopic parameters for $2 \times H_2$
rectangular configuration, as well its ground state energy.
\end{abstract}


\maketitle

\section{Motivation}
\label{sec:motivation}

The few-site models of correlated fermions play an important role in singling out, in an exact manner, the role of various local
intra-- and inter-site interactions against the hopping (i.e., containing both the covalent and the ionic factors) and thus, in establishing the optimal correlated state
of fermions\cite{Spalek-Oles,Penson,deBoer,Iglesias,Schumann, Schumann2, Acquarone3,Matlak} on local (nanoscopic) scale.
The model has also been used to obtain a realistic analytic estimate of the hydrogen-molecule energies of the ground and the excited states in
the correlated state \cite{Spalek}.
For this purpose, we have developed the so-called EDABI method, which combines \emph{E}xact \emph{D}iagonalization in the Fock space with a concomitant \emph{Ab} \emph{I}nitio
determination of the single-particle basis in the Hilbert space. So far, the method has been implemented by taking only $1s$ Slater orbitals, one per site \cite{Spalek-Gorlich}.
The method contains no parameters; the only approximation made is taking a truncated single-particle basis (i.e., one Slater orbital per site) when constructing the field
operator, that in turn is used to derive the starting Hamiltonian in the second-quantization representation. This Hamiltonian represents an extended Hubbard
Hamiltonian, with \emph{all} two-site interactions taken into account and the solution comprises not only the exact eigenvalues of the few-site Hamiltonian,
but also at the same time evaluation of the adjustable single-particle wave functions in the correlated state.
Also, the thermodynamic properties calculated rigorously exemplify \cite{Oles-Spalek,Spalek-Oles2} the low- and high-energy scales, corresponding to spin and local
charge fluctuations respectively. The former represents precursory magnetic-ordering effect whereas the latter local effects accompanying with the Mott-Hubbard transition.
In general, our approach follows the tradition of accounting for interelectronic correlations via the second-quantization procedure, with the adjustment of the single-particle
wave functions, contained in microscopic parameters of the starting model, in the correlated state.

The first aim of this paper is to extend a fully microscopic approach established earlier  \cite{Spalek, Spalek-Gorlich} and calculate all six possible
electron--ion coupling constants for $H_2$
as a function of the bond length. As a byproduct, we obtain an accurate estimate of the zero-point-motion amplitude and its energy to a high (ninth) order in
the relative displacement of the ions. This evaluation shows explicitly the dominant contributions to the vibronic spectra of the molecule. In effect,
the work formulates a complete two-site model of correlated states with all the coupling parameters calculated from an \emph{ab initio} procedure.
It also forms a starting point to a full scale dynamic calculations
involving a richer basis in the Hilbert space, at least in the adiabatic limit. So, although the importance of the present results to the discussion of $H_2$ molecule
exact evaluation of the ground-state energy
is limited, the approach may be extended to treat the molecular solid hydrogen with inclusion of interelectronic correlations. Explicitly, as a starting point
we calculate the intermolecular hopping integrals and the principal electron--electron interaction microscopic parameters as a function of intermolecular distance.

A methodological remark is in place here. As we determine the local ion--electron and electron--electron coupling parameters, they can be regarded as a starting estimate
of those for the bulk solid molecular hydrogen, as we have studied recently a critical pressure of metallization of the \emph{atomic} solid (Mott insulating) state \cite{Kadzielawa}.
The obtained pressure of atomic-hydrogen metallization is about $100 \ GPa$, the value which squares well with observed for the case of fluid molecular hydrogen ($140 \ GPa$)\cite{Weir},
although the recent simulations provide quite different values for fluid hydrogen analyzed at high temperature \cite{Sorella}.
Obviously, our previous work is not related directly to the molecular-hydrogen metallization in the solid state \cite{Ashcroft,Shibata,McMahon,Naumov}.
So far, we have discussed rigid-lattice properties. We believe that the present results form the first step in incorporating the vibrational spectrum and
correlations to extended systems.

The structure of the paper is as follows. Even though the main purpose of the paper is to calculate the local electron--proton and electron--electron coupling constants,
for the sake of the completeness, in Sections~\ref{sec:model}~\&~\ref{sec:molecule} we reproduce some of the results of \cite{Spalek} and correct some minor errors (cf. also
Appendix~\ref{app:solution}). In Sec.~\ref{sec:coupling} we define the method of calculation of both the electron--ion (proton) coupling constants (cf. also
Appendix~\ref{app:adiabatic}), as well as estimate the zero-point motion to the ninth order vs. the interionic distance.
In Sec.~\ref{sec:2molecule} we extend the single-molecule treatment and provide the intermolecular hopping amplitudes and the electron--electron microscopic parameters
which may serve for analysis of the solid molecular hydrogen.
Section~\ref{sec:results} contains physical discussion and a brief outlook, where we also refer to the finite-size Quantum Monte Carlo results.
In the series of Appendices we provide some analytical details, as they may form
analytical basis for the electron--lattice coupling supplementing the classic Slater results for the electronic part of $H_2$ molecule \cite{Slater}.

\section{Model and summary of purely electronic properties}
\label{sec:model}

\subsection{Wannier basis}
\label{ssec:basis}

To describe the behavior of an electron in the system of two ions we start from $1s$ Slater--type orbitals
\begin{align}
 \label{eq:slater}
 \Psi_i \left(  \vec{r} \right) = \sqrt{\frac{\alpha^3}{\pi}} e^{-\alpha | \vec{r} - \vec{R_i} | },
\end{align}
where $\alpha$ is the inverse size of the orbital.
To ensure orthogonality we use Wannier functions which in this case reduce the superposition of the atomic states, i.e.,
\begin{align}
 \label{eq:wannier}
 w_i \left(  \vec{r} \right) = \beta \Big[ \Psi_i \left(  \vec{r} \right) - \gamma \Psi_j \left(  \vec{r} \right) \Big],
\end{align}
with  the mixing parameters
\begin{equation}
\label{eq:mix}
\left\{
\begin{array}{l}
  \beta = \frac{1}{\sqrt{2}}\sqrt{\frac{1+\sqrt{1-S^2}}{1-S^2}} \\
  \gamma = \frac{S}{1+\sqrt{1-S^2}}
\end{array}
\right.
 \end{equation}
where $S = S (\alpha, R) \equiv \bracket{\Psi_1}{\Psi_2}$ is the atomic functions' overlap.

Eqs. \eqref{eq:mix} ensure both the orthogonality and proper behavior in atomic limit i.e., $\lim \limits_{R \rightarrow \infty} \beta = 1$, where $R$
is the average interatomic distance.
$\lim \limits_{R \rightarrow \infty} \gamma = 0$.

\subsection{Second-quantization picture}
\label{ssec:2quantization}

The two-site Hamiltonian with one orbital per site has the general form
\begin{align}
 \label{eq:ham}
  \mathcal{H} =& \epsilon \left( \NUM{1}{0} + \NUM{2}{0} \right) + t \sum_\sigma \left( \CR{1}{2} \AN{2}{2} +  \CR{2}{2} \AN{1}{2} \right)\\\notag
	      &+ U \left( \NUM{1}{1} \NUM{1}{-1} + \NUM{2}{1} \NUM{2}{-1} \right) - 2 J \vec{S}_1 \vec{S}_2 \\\notag
	      &+ \left( K - \frac{J}{2} \right) \NUM{1}{0} \NUM{2}{0} + J \left( \CR{1}{1} \CR{1}{-1} \AN{2}{-1} \AN{2}{1} + h.c. \right)\\\notag
	      &+V \sum_\sigma \left[ \left( \NUM{1}{2} + \NUM{2}{2} \right) \left( \CR{1}{-2}\AN{2}{-2} + \CR{2}{-2}\AN{1}{-2}  \right) \right],
\end{align}
where $\AN{i}{2}$ and $\CR{i}{2}$ are the fermionic operators of annihilation and creation of the electron with spin $\sigma$ on $1s$ orbital of hydrogen atom
$i \in \{ 1 , 2 \}$.

The microscopic parameters $\epsilon = T_{11}$, $t=T_{12}$, $U=V_{1111}$, $J=V_{1122}$, $K=V_{1212}$ and $V=V_{1112}$ correspond to one-- and two-particle interactions \cite{Spalek}
\begin{subequations}
 \label{eq:microDefs}
 \begin{align}
  \label{eq:microDefsT}
    T_{ij} &= \matrixel{w_i}{\mathcal{T}}{w_j}, \\
  \label{eq:microDefsV}
    V_{ijkl} &= \matrixel{w_i w_j}{\mathcal{V}}{w_k w_l},
 \end{align}
\end{subequations}
where in atomic units $\mathcal{T} = - \bigtriangledown^2 - {2} / | \vec{r} - \vec{R} |$, and $\mathcal{V} =  {2} / | \vec{r} - \vec{r}' |$.
In the Appendix~\ref{app:solution} we provide explicitly the form of microscopic parameters as a function of both intersite static distance $R$ and
the inverse wave-function size $\alpha$. In what follows we diagonalize first \eqref{eq:ham}, and subsequently optimize the wave functions
contained in the microscopic parameters of \eqref{eq:ham}. This program will be carried out systematically in what follows.

\subsection{Exact solution}
\label{ssec:solution}

System described by the Hamiltonian \eqref{eq:ham} has an exact solution previously studied in detailed in \cite{Spalek}.
For two-electron system ($n_1 +n_2=2$), i.e. with one particle per site, the starting basis is 
\begin{subequations}
\label{eq:vecs}
 \begin{align}
\label{eq:vec1}
  \ket{1} &= \CR{1}{1}\CR{2}{1} \ket{0}, \\
\label{eq:vec2}
  \ket{2} &= \CR{1}{-1}\CR{2}{-1} \ket{0}, \\
\label{eq:vec3}
  \ket{3} &= \frac{1}{\sqrt{2}} \left( \CR{1}{1}\CR{2}{-1} + \CR{1}{-1}\CR{2}{1}  \right) \ket{0},
\end{align}
representing the intersite spin-triplet states with eigenvalues $E_1=E_2=E_3=2 \epsilon + K - J$, and
\begin{align}
\label{eq:vec4}
  \ket{4} &= \frac{1}{\sqrt{2}} \left( \CR{1}{1}\CR{2}{-1} - \CR{1}{-1}\CR{2}{1}  \right) \ket{0}, \\
\label{eq:vec5}
  \ket{5} &= \frac{1}{\sqrt{2}} \left( \CR{1}{1}\CR{1}{-1} + \CR{2}{1}\CR{2}{-1}  \right) \ket{0}, \\
\label{eq:vec6}
  \ket{6} &= \frac{1}{\sqrt{2}} \left( \CR{1}{1}\CR{1}{-1} - \CR{2}{1}\CR{2}{-1}  \right) \ket{0},
 \end{align}
\end{subequations}
representing the spin-singlet states, with the corresponding Hamiltonian matrix involving the matrix elements
$\matrixel{i}{\mathcal{H}}{j} \equiv \left( \mathcal{H}_{ij} \right)$:
\begin{equation}
\label{eq:hamMat}
 \left( \mathcal{H}_{ij} \right) = 
\left(
 \begin{array}{ccc}
2 \epsilon + K + J & 2(t+V) &0 \\
2(t+V) & 2 \epsilon + U + J &0 \\
0&0&2 \epsilon + U - J
\end{array}
\right).
\end{equation}

The state \eqref{eq:vec6} is an eigenvector of \eqref{eq:hamMat} with eigenvalue $E_6 = 2 \epsilon + U - J$. The diagonalization
supplies us with the two other eigenvectors \cite{Spalek-Oles}
 \begin{align}
  \label{eq:vecSol}
   \ket{\pm} = [ 2 \mathcal{D} ( \mathcal{D} \mp U &\pm K  )] ^{-\frac{1}{2}}\\\notag
  &\Big[ \mp ( \mathcal{D} \mp U \pm K ) \ket{4} + 4 | t + V | \ket{5} \Big],
 \end{align}
with eigenvalues
\begin{align}
 \label{eq:EigE}
  E_\pm = 2 \epsilon + \frac{U+K}{2} + J \pm \frac{1}{2} \mathcal{D},
\end{align}

where $\mathcal{D} = \sqrt{ \left( U - K \right)^2 + 16 \left( t + V \right)^2}$. The state $\ket{-}$
from \eqref{eq:vecSol} is the lowest-energy spin-singlet eigenstate. It is this state, for which we determine explicitly the single-particle
wave function and subsequently, determine the microscopic parameters $\epsilon$, $t$, $U$, $J$, $K$, and $V$ explicitly, all as a function of
interionic distance $R$.

\subsection{Optimization of the atomic basis}
\label{ssec:optimization}

The ground-state energy is the energy $E_-$ of \eqref{eq:EigE}, supplemented with the ion--ion repulsion, i.e. by
\begin{align}
 \label{eq:Eg}
  E_G = E_- + \frac{2}{R},
\end{align}
where $(2/R)$ is represented also in atomic units. As all the microscopic parameters are only a function of the distance $R$
and the inverse wave-function size $\alpha$, we have $E_G = E_G \left( \alpha, R \right)$.
For each distance $R$, we minimize $E_G$ with respect to $\alpha$, thus closing the solution. Finally, we select $R=R_B$ as the equilibrium solution,
for which still the zero-point motion has to be taken into account.

\section{Stationary state for $H_2$ molecule}
\label{sec:molecule}

\begin{figure}
\centering
\includegraphics[width=\figuresize]{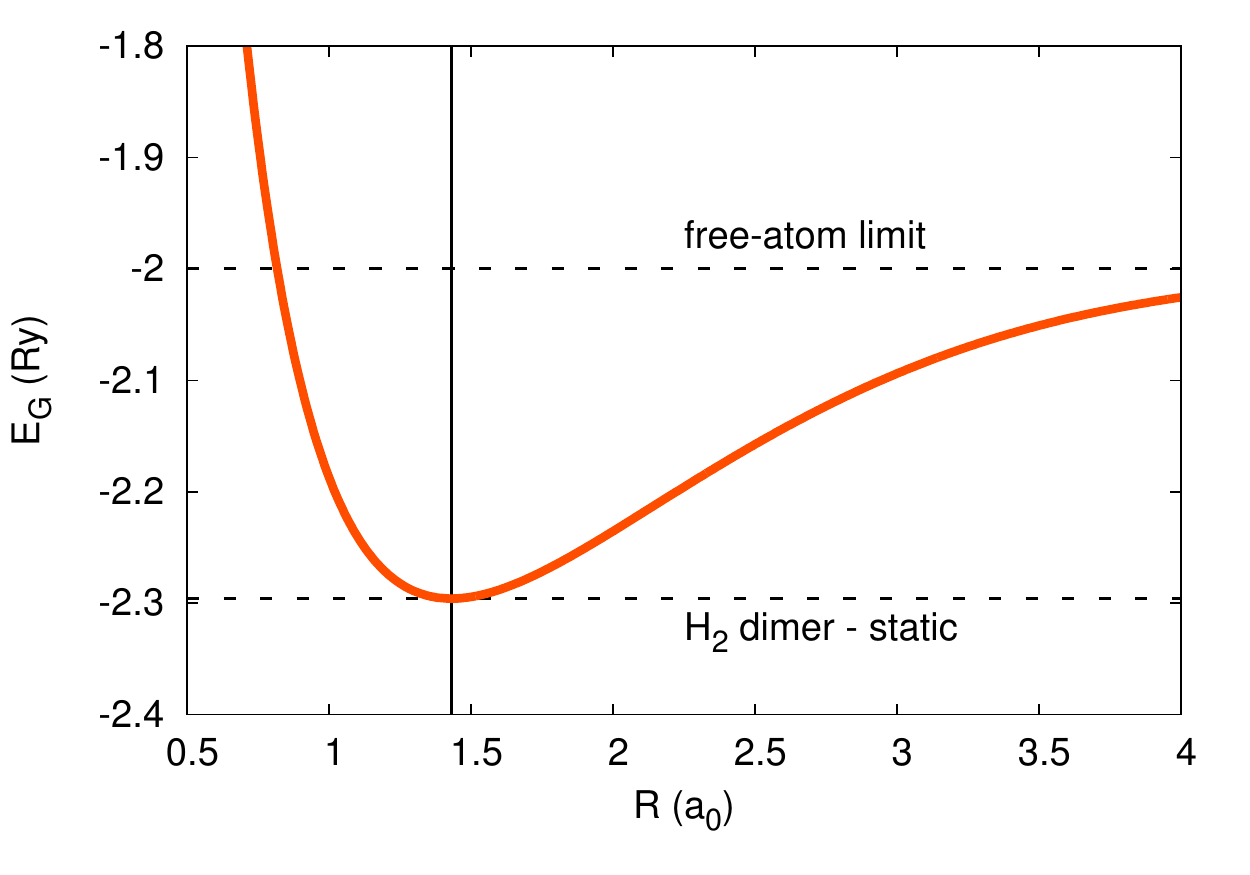}
\caption{Ground-state energy -- as defined via \eqref{eq:Eg} -- versus interionic static distance $R$. Note that the minimum value is 
$E_B = -2.29587 \ Ry$ (marked by the vertical line here and below) at $R_B = 1.43042 \ a_0$.}
\label{fig:energy}
\end{figure}

\begin{figure}
\centering
\includegraphics[width=\figuresize]{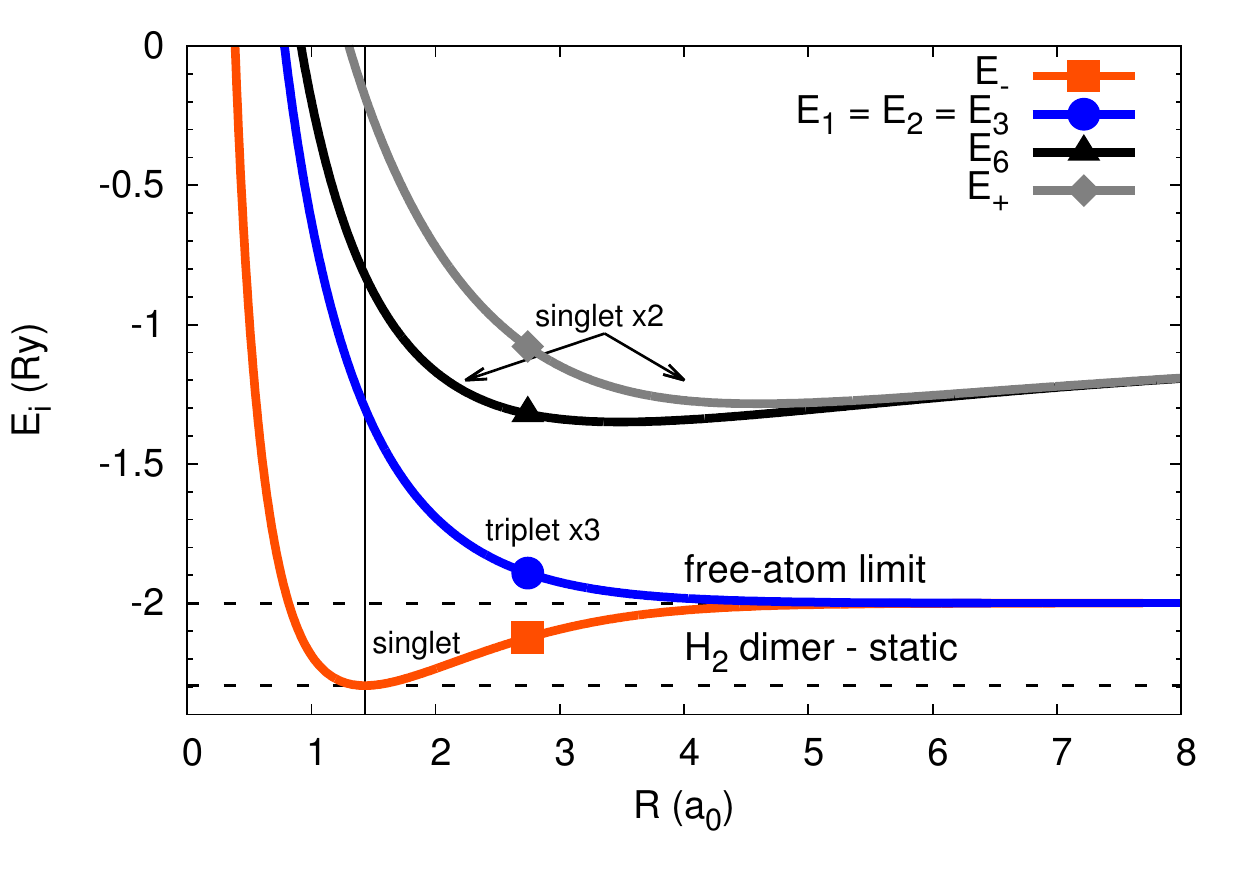}
\caption{Solutions for the states: the spin-triplets ($E_1=E_2=E_3$) and the spin-singlets ($E_\pm$, $E_6$) versus the interionic distance $R$.
The spin-singlet state $\ket{-}$ is the equilibrium state.}
\label{fig:energies}
\end{figure}

\begin{figure}
\centering
\includegraphics[width=\figuresize]{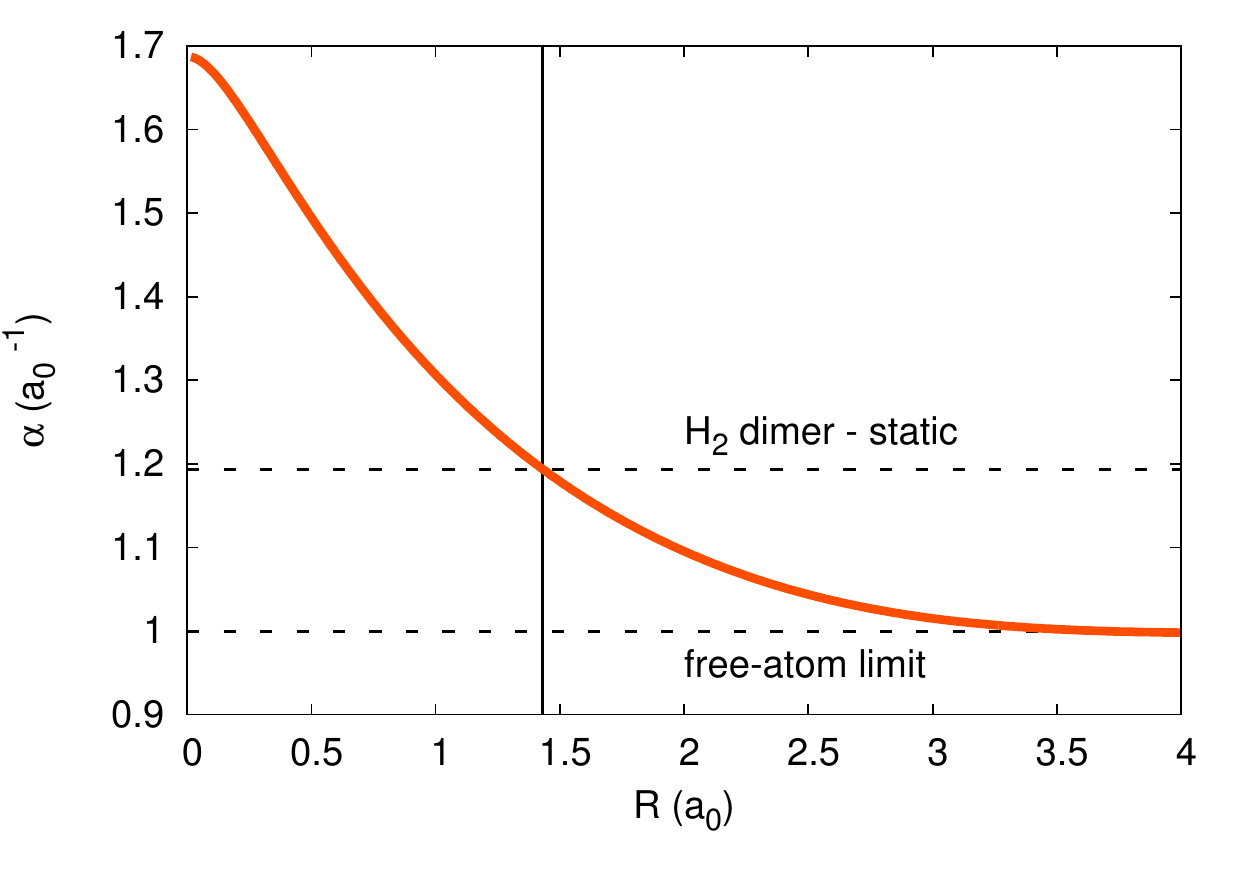}
\caption{The optimal inverse wave-function size $\alpha$ versus the proton--proton average distance $R$. Note that $\alpha_B = 1.19378 a_0 ^{-1}$.}
\label{fig:alpha}
\end{figure}

In Fig.~\ref{fig:energy} we plot the energy of $H_2$ (dimer) versus the distance $R$. It is crucial that we obtain a local (and global) minimum
at~$R=R_B \equiv 0.757 \Ang$. This simple result, obtained in \cite{Spalek} differs with respect to the virtually exact solution by Ko\l{}os and Wolniewicz \cite{Kolos,Szabo},
$R_{K-W} = 0.74 \Ang$ by $2.5 \%$ only.

\begin{figure}
\centering
\includegraphics[width=\figuresize]{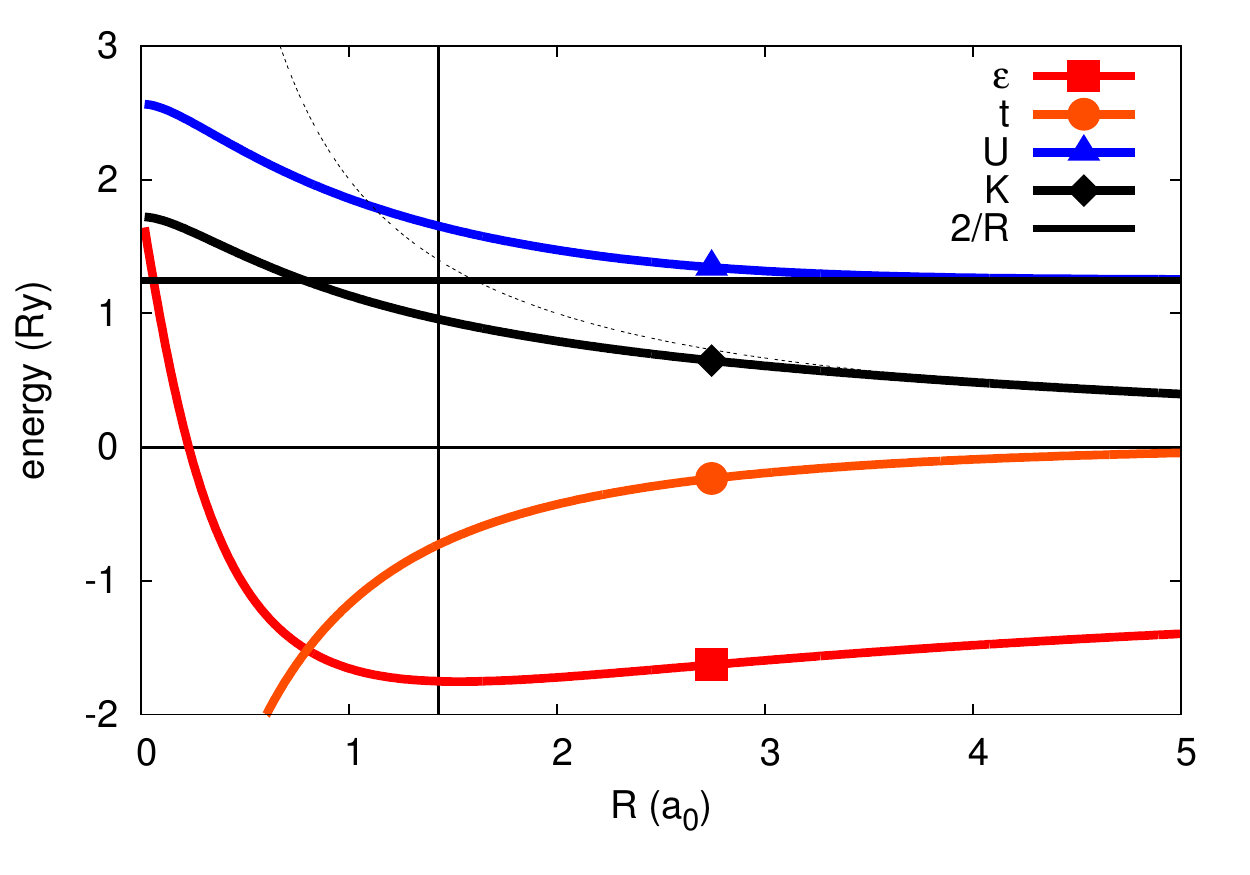}
\caption{Microscopic parameters $\epsilon$, $t$, $U$, and $K$ versus average interionic distance $R$. Note the convergence of the intersite Coulomb repulsion $K$ to
the classical value $2/R$ (dashed line) at $R \rightarrow \infty$. The on-site repulsion $U$ reaches also its atomic limit $U_{at} = 1.25 \ Ry$, whereas
the hopping parameter $t \rightarrow 0$.}
\label{fig:microscopic}
\end{figure}

\begin{figure}
\centering
\includegraphics[width=\figuresize]{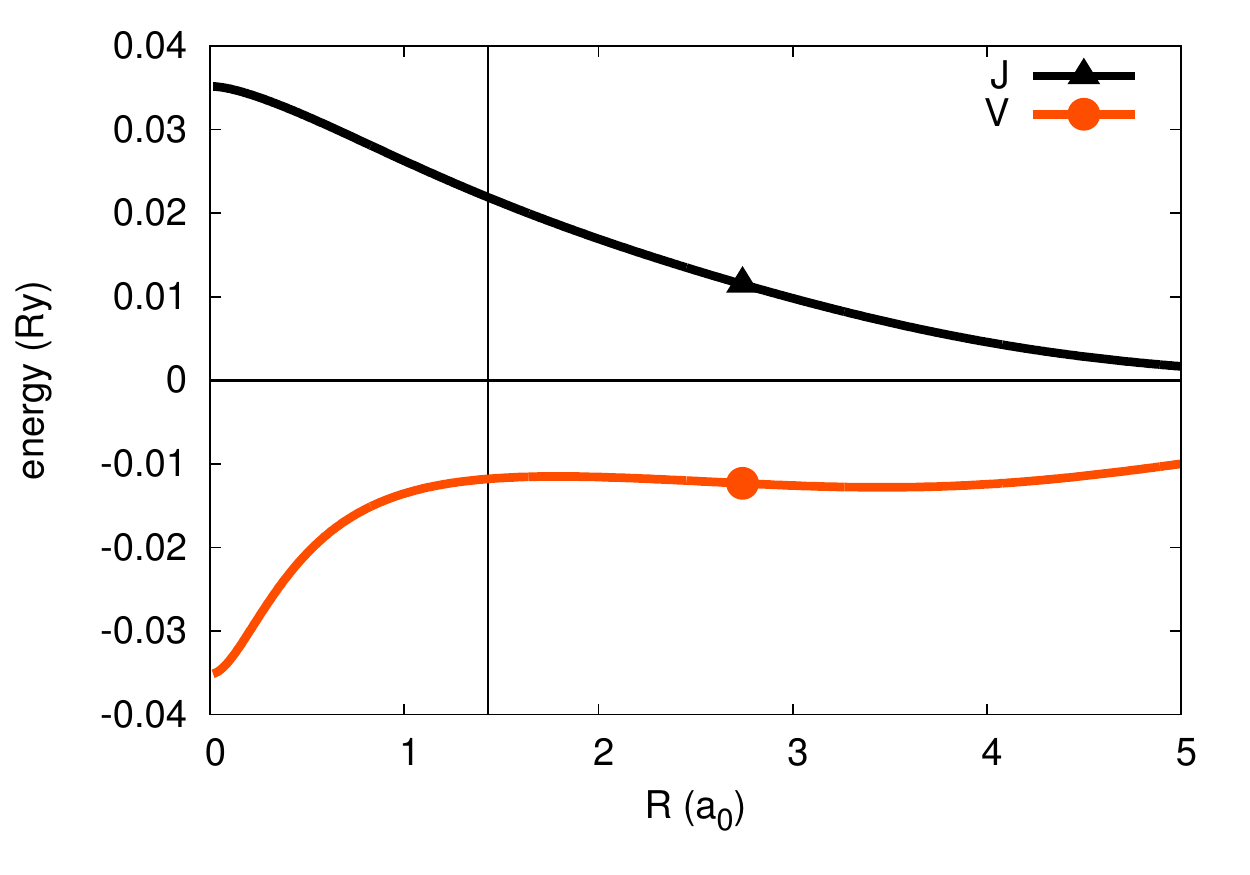}
\caption{Microscopic parameters $J$ and $V$ vs. $R$. Note that the exchange integral is always ferromagnetic, and the so-called correlated
hopping parameter is $V<0$.}
\label{fig:microscopic2}
\end{figure}

In Fig.~\ref{fig:energies} we plot the sequence of the spin-singlets and the spin-triplet states. Parenthetically, the start from second-quantization language allows for
evaluation of the ground-state and the lowest excited states, on an equal footing. This feature provides the difference with purely variational calculations
in~the~first-quantization language. Namely, within this basis the spin-singlet state is stable at arbitrary interionic distance $R$. In Figs.~\ref{fig:alpha} - ~\ref{fig:microscopic2} the inverse wave function
size $\alpha$, as well as all the microscopic parameters, are all displayed as a function of the bond length $R$. One can see that with the increasing $R$ values tends to the proper free-atom limits. Those
quantities form an input for the subsequent evaluation of electron-proton coupling constants discussed next.

\section{Adiabatic approximation for the electron--ion coupling}
\label{sec:coupling}

Our principal aim here is to extend our previous model \cite{Spalek, Spalek-Gorlich} by allowing the ions oscillate around the equilibrium positions.
Thus the interionic distance $R$ is taken now in the form
\begin{align}
 \label{eq:sizeR}
 R = R_B + \delta R,
\end{align}
where $ \delta R$ is responsible for the zero-point motion. The electronic part of the ground-state energy is expanded next on $\delta R$ in terms
of~a~Taylor series, which to the ninth order reads
\begin{align}
 \average{\mathcal{H}}_{\delta R} = E_B + \sum_{i=2}^9  \frac{1}{i !} E_B^{(i)}\delta R^i + O \left( \delta R ^{10} \right),
\end{align}
where $E_B^{(i)} =\left. \frac{\partial^{i} E_B}{\partial R ^i}\right|_{R_B}$ and $E_B^{(1)} = 0$, whereas all the remaining terms but
for the energy $E_B$ describe the oscillations (cf. Table~\ref{tab:energies} for numerical values).
We have modified the Hamiltonian \eqref{eq:ham} accordingly by taking into account $\delta R$, i.e.,
\begin{align}
\label{eq:newHam}
 \mathcal{H} \rightarrow \mathcal{H} + \delta \mathcal{H},
\end{align}

where $\delta \mathcal{H}$ is the additional term. Also, $\mathcal{H}$ simplifies to the form
\begin{align}
 \mathcal{H} = \sum_i \Xi_i \hat{O}_i,
\end{align}
where $\Xi = \{ \epsilon,\ t,\ U,\ J,\ K,\ V \}$ and $\hat{O}_i$ are the corresponding operator parts of Hamiltonian: the two- and four-operator terms of \eqref{eq:ham}
standing next to the respective microscopic parameter (for example $\hat{O}_{\epsilon} = \hat{n}_1 + \hat{n}_2$). With the Hamiltonian in this form we now have
the energy change due to the change of the microscopic parameters

\begin{align}
 \label{eq:coupling}
 \delta \mathcal{H} = \sum_i \delta \Xi_i \hat{O}_i =  \sum_i \xi_i \delta R \ \hat{O}_i,
\end{align}
where $\xi_i \equiv \frac{\delta \Xi_i}{\delta R}$. Since $\delta R \propto ( b_i ^{\dagger} + b_i ^{} )$, where $b_i ^{\dagger}$, $b_i ^{}$ are bosonic creation and annihilation
operators of the system deformation and the set $\{ \xi_i \}$ defines a new set of microscopic parameters - the electron--ion coupling constants.
They can be derived by differentiation in~a~way similar to that of \cite{Acquarone,Spalek2} (cf. Appendix~\ref{app:adiabatic} for details).

\begin{figure}
\centering
\includegraphics[width=\figuresize]{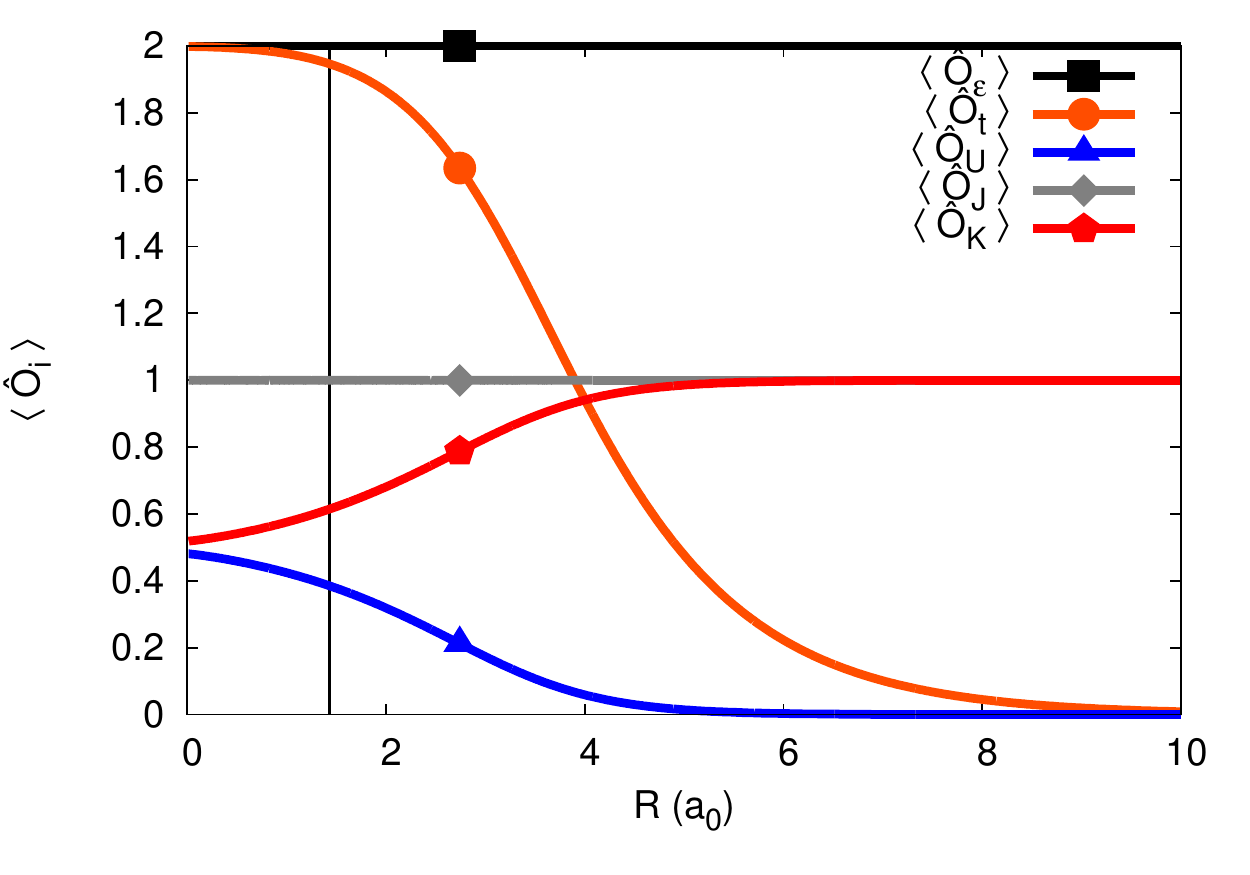}
\caption{Averages \eqref{eq:avs} calculated in the ground-state versus distance $R$. They are of the order of unity.}
\label{fig:averages}
\end{figure}

\begin{figure}
\centering
\includegraphics[width=\figuresize]{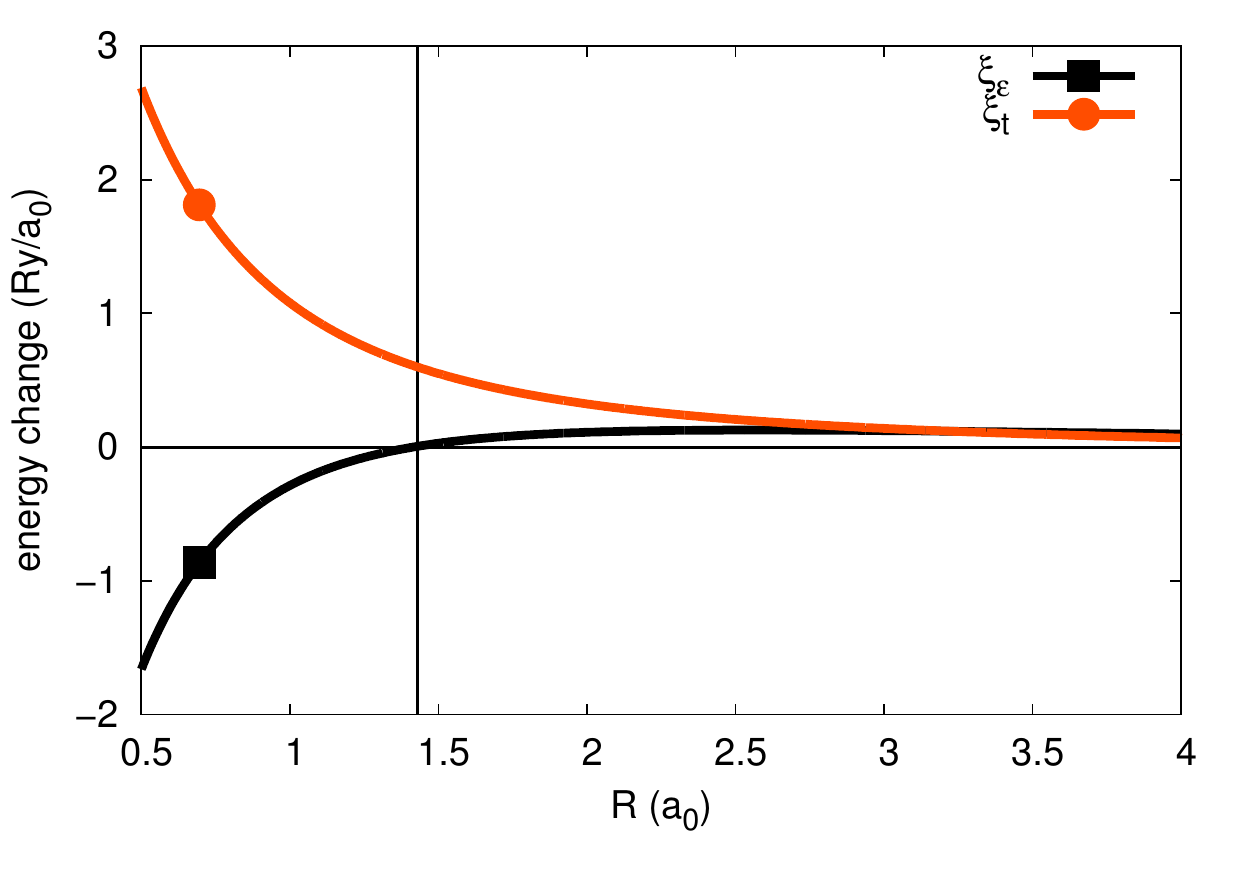}
\caption{Coupling constants $\xi _\epsilon$ and $\xi_t$  versus intersite distance $R$.}
\label{fig:couplings1}
\end{figure}

\begin{figure}
\centering
\includegraphics[width=\figuresize]{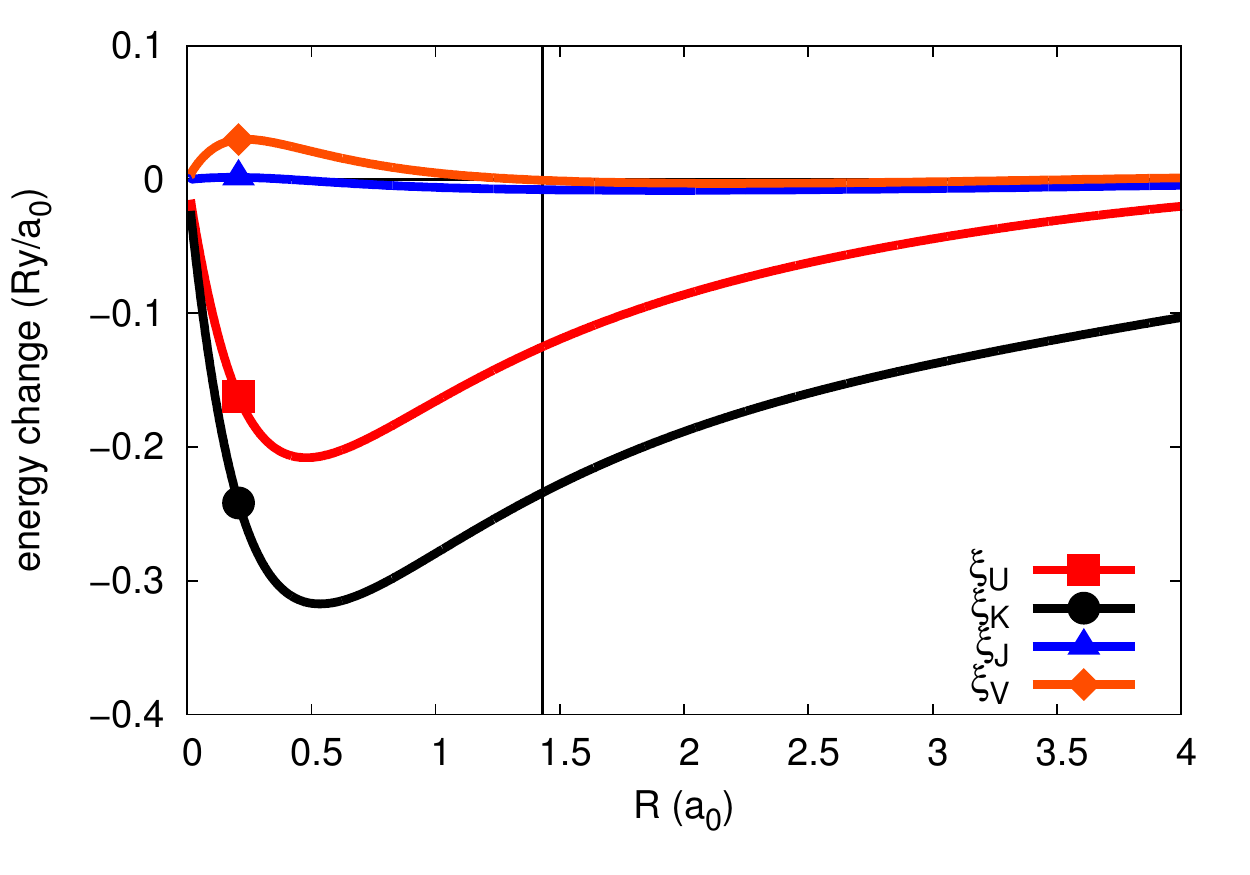}
\caption{Coupling constants $\xi _U$, $\xi _K$, $\xi _J$ and $\xi_V$  versus intersite distance $R$}
\label{fig:couplings2}
\end{figure}

The shift of the ions changes the system properties in the following manner
\begin{align}
 \average{\delta \mathcal{H}} =  \sum_i \xi_i \average{\hat{O}_i}_{0} \delta R,
\end{align}
where the average $\average{\hat{O}_i}_{0} = \matrixel{-}{\hat{O}_i}{-}$ is taken with respect to the ground state. In effect, we obtain
\begin{subequations}
\label{eq:avs}
\begin{align}
\label{eq:avse}
 \average{\hat{O}_\epsilon}_{0} &= 2,\\
\label{eq:avst}
 \average{\hat{O}_t}_{0} &= \frac{8 |t + V|}{\mathcal{D}},\\
\label{eq:avsU}
 \average{\hat{O}_U}_{0} &= \frac{16 (t + V)^2}{2\mathcal{D}(\mathcal{D}+U-K)},\\
\label{eq:avsJ}
 \average{\hat{O}_J}_{0} &= 1,\\
\label{eq:avsK}
 \average{\hat{O}_K}_{0} &= \frac{\mathcal{D} + U - K}{2\mathcal{D}},\\
\label{eq:avsV}
 \average{\hat{O}_V}_{0} &= \frac{8 |t + V|}{\mathcal{D}}.
\end{align}
\end{subequations}

The $R$ dependence of the parameters $\average{\hat{O}_i}$ given by \eqref{eq:avs} is displayed in Fig.~\ref{fig:averages}. As they are of the order of unity,
the principal factor determining the relative strength of the coupling constants are provided by the parameters $\{ \xi_i \}$ displayed in Figs.~\ref{fig:couplings1}
and \ref{fig:couplings2}. At the equilibrium bond distance marked by the vertical line, the largest values are (on the absolute scale) those coming from modulation
of the hopping parameter ($\xi_t$) and the change of intersite Coulomb interaction ($\xi_K$). The first of the two has been included in Su, Schrieffer and Heeger
model \cite{Su}. The second may play an important role in the high-$T_C$ superconductivity \cite{Spalek2}. Also, we see that the so-called Holstein
coupling \cite{Holstein} is not important if calculated near the hydrogen-molecule equilibrium state.

We determine the value of $\delta R$ by minimizing the total energy of the system:
\begin{align}
 \label{eq:Etotal}
 E_{total} \equiv &\average{\mathcal{H}} + \average{\delta \mathcal{H}} + \average{\mathcal{H}_{ion}}. \\\notag
\end{align}
where
\begin{align}
 \label{eq:ionham}
\average{\mathcal{H}_{ion}} = 2\frac{\delta P ^2}{2 M} + \frac{2}{R + \delta R}.
\end{align}
where the ionic momentum $\delta P$ is evaluated via Heisenberg principle and $M \stackrel{a.u.}{\approx} 1836.15267 m_e$ is the mass of the proton.

\section{Evaluation of the microscopic parameters for the two-molecule system}
\label{sec:2molecule}

\begin{figure}
\centering
\includegraphics[width=\figuresize]{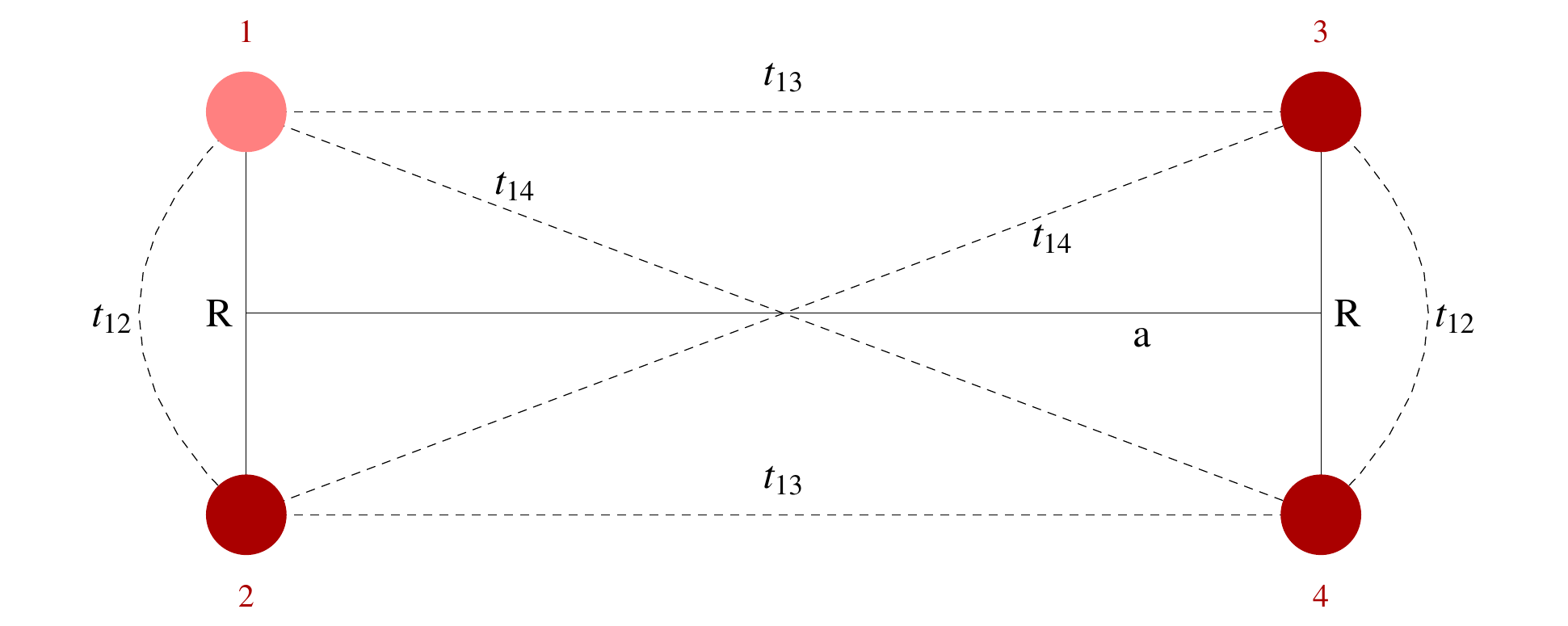}
\caption{The system of two $H_2$ molecules at the relative distance $a$. The hopping integrals $t_i$ are marked next to the respective dashed lines. Note that the orthogonalization
procedure for four sites produces a different basis than that obtained in \eqref{eq:mix}.}
\label{fig:4atoms}
\end{figure}

We extend our approach by considering a system of two $H_2$ dimers at the relative distance $a$ from each other (cf. Fig.~\ref{fig:4atoms}). We calculate the respective
hopping integrals, where $t_{12}$ should approach $t$ defined in \eqref{eq:texpr} for large $a$, as well as the new single-particle energy $\epsilon$ should again
converge to previously obtained value \eqref{eq:eexpr}. We determine all the two-particle interaction integrals, thus going beyond the Hubbard model solved in \cite{Schumann}.
Additionally, in Table~\ref{tab:2H2} we list the numerical values of the most relevant microscopic parameters.

\begin{figure}
\centering
\includegraphics[width=\figuresize]{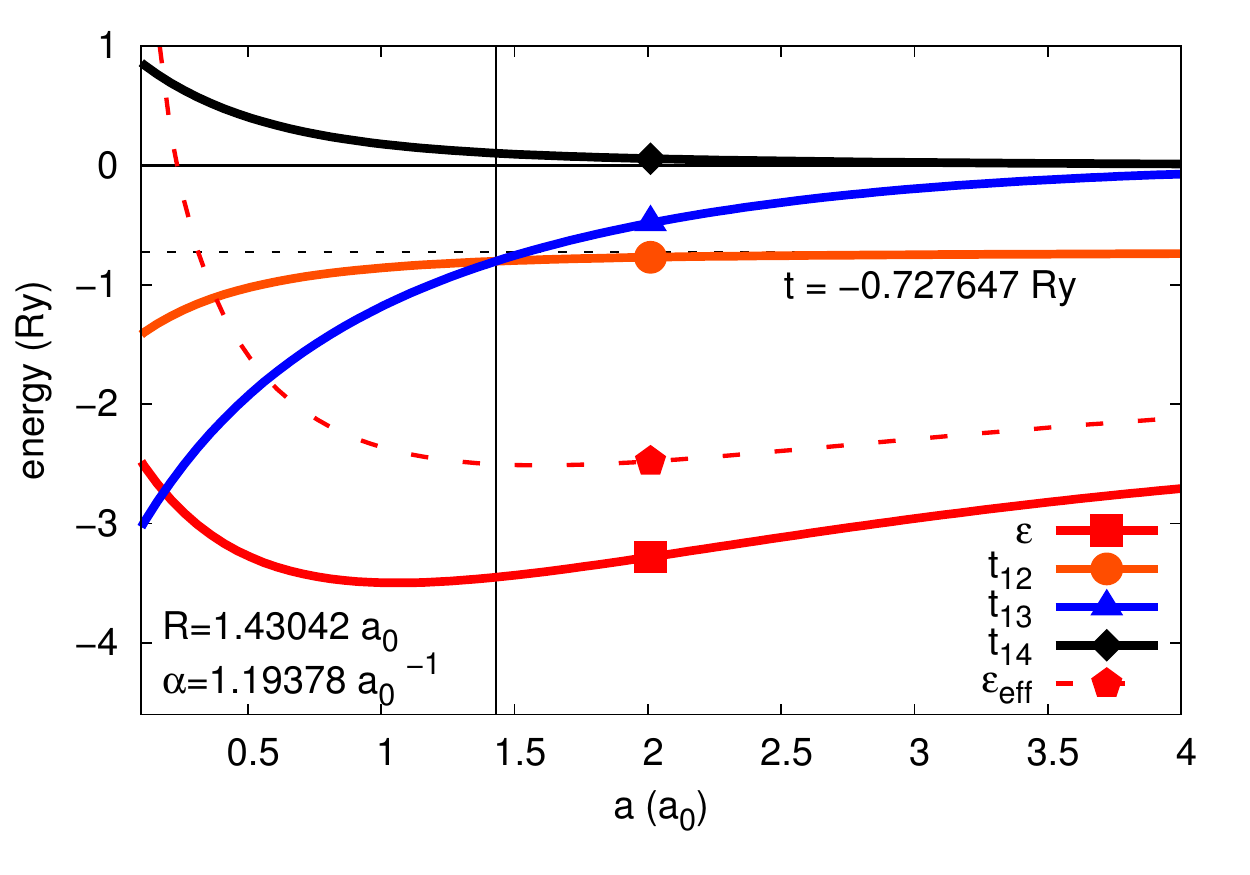}
\caption{The one-particle microscopic parameters for two $H_2$ molecules system vs. intermolecular distance $a$. The red dashed line marks
the effective (renormalized by ion--ion repulsion) single-particle
energy per site $\epsilon_{eff} = \epsilon + 1/N \sum_i 2/R_i$.
Note the convergence of $t_{12} \rightarrow t$, and $t_{13}, \ t_{14} \rightarrow 0$ with $a \rightarrow \infty$. The equality of $t_{12}$
and $t_{13}$ at $a=R_B=1.43042 \ a_0$ should be observed as well.}
\label{fig:2H2et}
\end{figure}

\begin{figure}
\centering
\includegraphics[width=\figuresize]{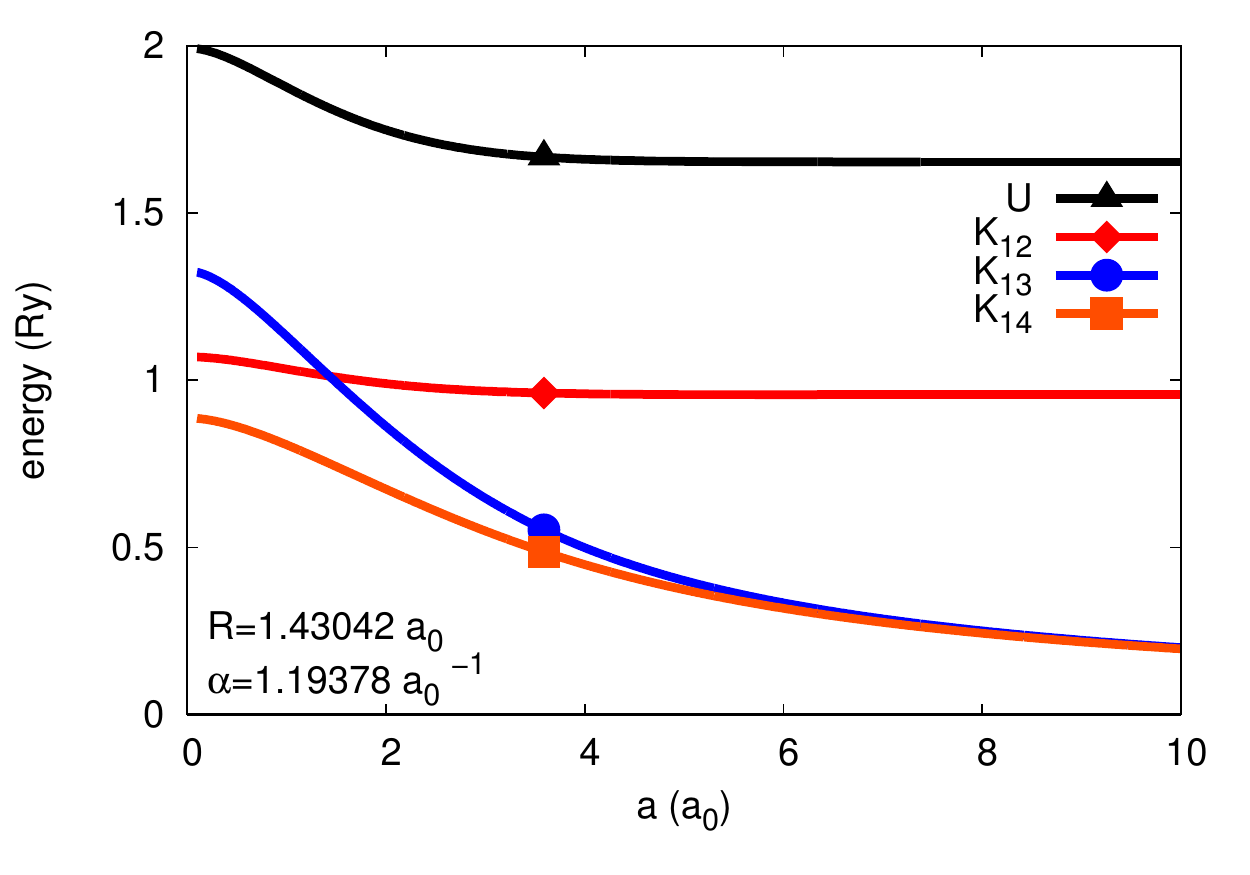}
\caption{
Coulomb-interaction microscopic parameters: the on-site part ($U$), intramolecular ($K_{12}$), and intermolecular $K_{13}$ and $K_{14}$ for
two $H_2$-molecule system vs. intermolecular distance $a$.
}
\label{fig:2H2UK}
\end{figure}

\begin{figure}
\centering
\includegraphics[width=\figuresize]{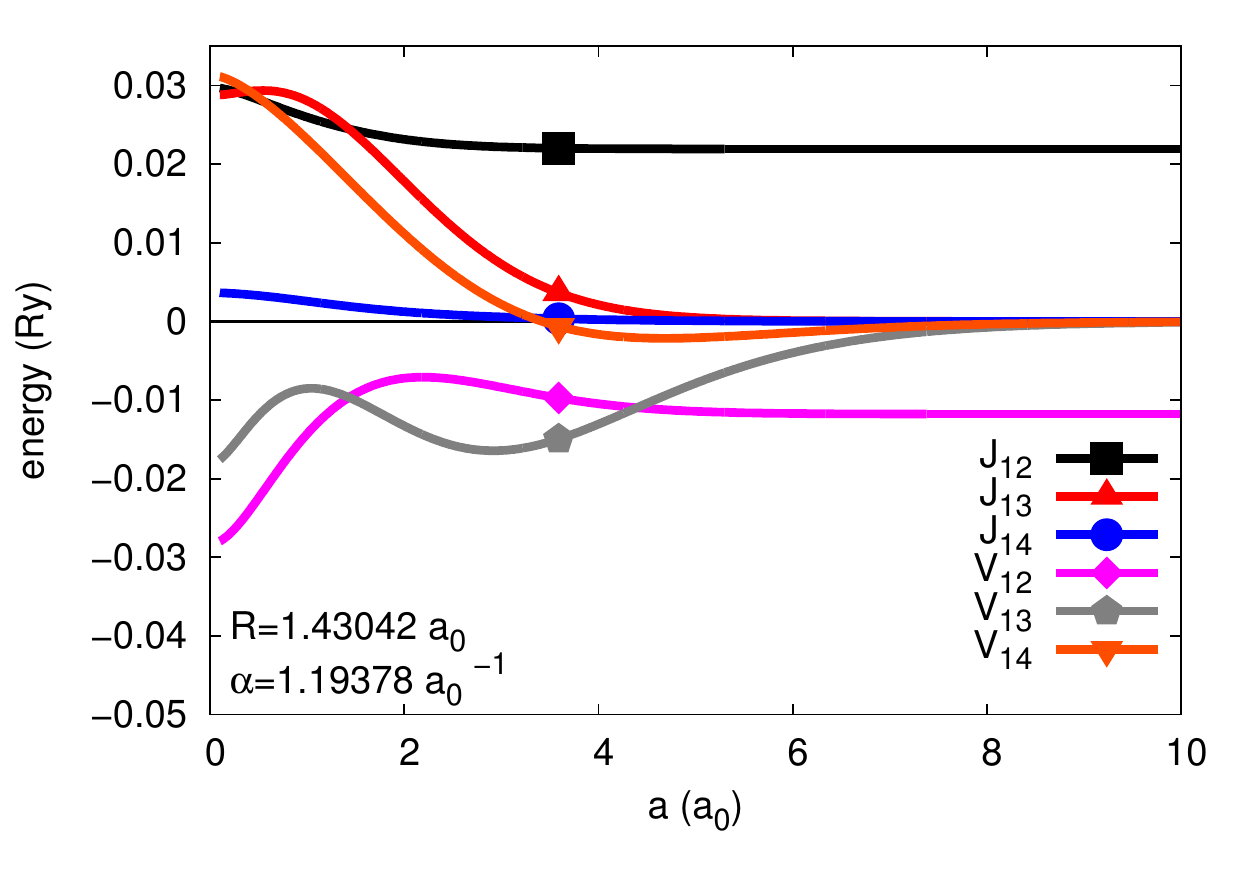}
\caption{
Two-particle microscopic parameters: intramolecular spin-exchange $J_{12}$ and correlated hopping $V_{12}$,
as well as the intermolecular parameters $J_{13}$, $J_{14}$, $V_{13}$ and $V_{14}$ for two $H_2$ molecules system vs. intermolecular distance $a$. Note that all the intermolecular
parameters converge to zero quickly.
}
\label{fig:2H2VJ}
\end{figure}

\begin{figure}
\centering
\includegraphics[width=\figuresize]{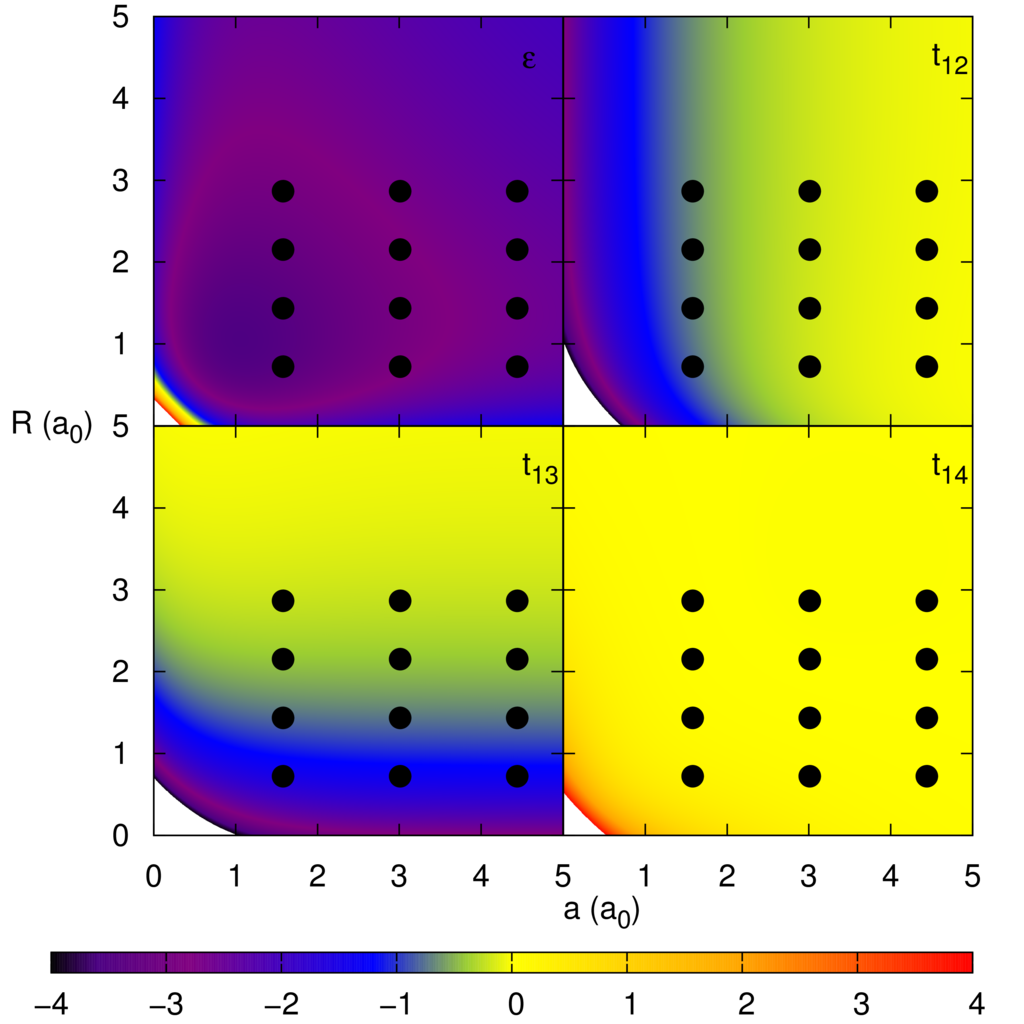}
\caption{The one-particle microscopic parameters (in $Ry$) for two $H_2$ molecules system vs. intermolecular distance $a$ and interionic distance $R$. Note the symmetry
of $\epsilon$ and $t_{14}$. As expected for relatively small distances values of $t_{12}$ and $t_{13}$ are negative, whereas $t_{14}$ is positive. When approaching point $(0,0)$ all
parameters diverge to minus ($t_{12}$ and $t_{13}$) or plus ($\epsilon$ and $t_{14}$) infinity. The explicit values of the marked points are given in Tab.~\ref{tab:2H2}}
\label{fig:map}
\end{figure}

The results are presented in Figs.~\ref{fig:2H2et}, \ref{fig:2H2UK}, \ref{fig:2H2VJ} and \ref{fig:map}. Note that all the results converge to the free-molecule ($a \rightarrow \infty$)
values. The calculated hopping values of $t_{13}$ and $t_{14}$ may serve as input parameters for $H_2$ molecular crystal.

\begin{table*}
\caption{Numerical values of single-particle energy ($\epsilon$), the hopping integrals ($t_{\alpha \beta}$),
the on-site Coulomb repulsion $U$ and the intramolecular Coulomb interaction $K_{12}$ for the two-molecule system, values refer to the points marked in the Fig.~\ref{fig:map}.}
\label{tab:2H2}
\begin{tabular}{||d|d||d||d|d|d||d||d|d|d||}\hline
\mc{1}{||c|}{$R \ (a_0)$} & \mc{1}{c||}{$a \ (a_0)$} & \mc{1}{c||}{$\epsilon \ (Ry)$} & \mc{1}{c|}{$t_{12} \ (Ry)$} & \mc{1}{c|}{$t_{13} \ (Ry)$} & \mc{1}{c||}{$t_{14} \ (Ry)$} & \mc{1}{c||}{$U \ (Ry)$} & \mc{1}{c|}{$K_{12} \ (Ry)$} & \mc{1}{c|}{$K_{13} \ (Ry)$} & \mc{1}{c||}{$K_{14} \ (Ry)$}\\\hline
0.715 & 1.43 & -3.4265 & -1.5534 & -0.9320 & 0.2799 & 1.9210 & 1.2082 & 1.0480 & 0.8405\\\hline 
0.715 & 2.86 & -2.9068 & -1.3948 & -0.2349 & 0.0993 & 1.7875 & 1.1386 & 0.6790 & 0.5976\\\hline 
0.715 & 4.29 & -2.5229 & -1.3671 & -0.0499 & 0.0339 & 1.7557 & 1.1233 & 0.4674 & 0.4369\\\hline 

1.43 & 1.43 & -3.4500 & -0.8030 & -0.8030 & 0.1023 & 1.8143 & 1.0127 & 1.0127 & 0.7514\\\hline 
1.43 & 2.86 & -3.0007 & -0.7504 & -0.2232 & 0.0279 & 1.6903 & 0.9699 & 0.6732 & 0.5655\\\hline 
1.43 & 4.29 & -2.6483 & -0.7380 & -0.0535 & 0.0096 & 1.6585 & 0.9587 & 0.4666 & 0.4245\\\hline 

2.145 & 1.43 & -3.2344 & -0.4278 & -0.7651 & 0.0500 & 1.7359 & 0.8269 & 0.9858 & 0.6552\\\hline 
2.145 & 2.86 & -2.8805 & -0.4211 & -0.2294 & 0.0013 & 1.6162 & 0.8047 & 0.6674 & 0.5223\\\hline 
2.145 & 4.29 & -2.5668 & -0.4185 & -0.0610 & -0.0019 & 1.5839 & 0.7977 & 0.4655 & 0.4056\\\hline 

2.86 & 1.43 & -3.0007 & -0.2232 & -0.7504 & 0.0279 & 1.6903 & 0.6732 & 0.9699 & 0.5655\\\hline 
2.86 & 2.86 & -2.7193 & -0.2354 & -0.2354 & -0.0075 & 1.5712 & 0.6631 & 0.6631 & 0.4739\\\hline 
2.86 & 4.29 & -2.4410 & -0.2385 & -0.0670 & -0.0066 & 1.5383 & 0.6593 & 0.4646 & 0.3816\\\hline 

\end{tabular}
\end{table*}

Explicitly, in Fig.~\ref{fig:2H2et} we display the intermolecular dependence of single-particle parameters. For the distances $a \gtrsim 2 \ a_0$ the hoppings
$t_{13}$ and $t_{14}$ can be regarded as small on the scale $t_{12}=t$. Hence, the system in solid will preserve its molecular character, with no magnetism
involved even though we have nominally one electron per atom. In other words, the lowest band will be full and no simple-minded Hubbard subband (HOMO-LUMO) picture
in the ground state appears. In Fig.~\ref{fig:2H2UK} we compare the relative values of intramolecular ($U$, $K_{12}=K$) versus intermolecular ($K_{13}$, $K_{14}$)
Coulomb interactions. Again, intramolecular interactions dominate for $a \gtrsim 2 \ a_0$. From Figs.~\ref{fig:2H2et} and \ref{fig:2H2UK} it follows then, that in
the insulating (molecular-crystal) state virtual hopping processes in a similar manner to the kinetic exchange, will contribute and renormalize the gap between
the full-band (valence) and the conduction-band (excited single electron) states. This gap will have the form of the Hubbard gap, as the value of $U$,
corresponding to the transition $2H_2 \rightarrow H_2^- + H_2^+$ will have the value $U-K_{12} \approx 0.6 \ Ry$, by far the largest energy in
the insulating state. For the sake of completeness, we have plotted in Fig.~\ref{fig:2H2VJ} the remaining interaction parameters: the exchange
integrals, intra- ($J_{12}$) and inter-molecular ($J_{13}$ and $J_{14}$), as well as the correlated hopping amplitudes: $V_{12}$ and ($V_{13}$ and $V_{14}$),
respectively.

\begin{figure}
\centering
\includegraphics[width=\figuresize]{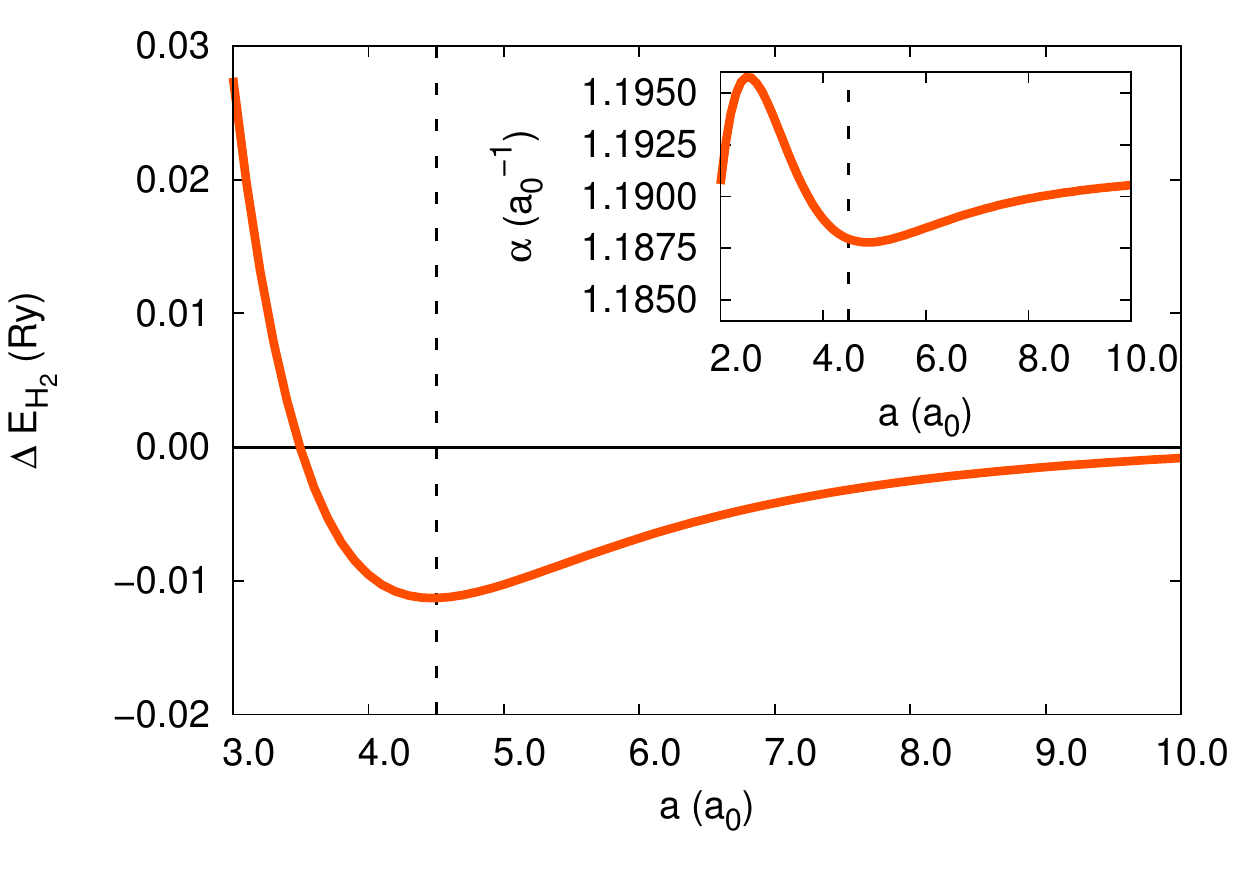}
 \caption{Difference between the energies of the $\left( H_2 \right)_2$ system and that of the free molecules (per molecule) versus intermolecular distance $a$. Note the
 van der Waals-like behavior with the shallow minimum at $a=4.5 \ a_0$.
 Inset: inverse atomic wave-function size $\alpha$ versus $a$. For $a \rightarrow \infty$, it approaches the value $\alpha_B = 1.19378 \ a_0 ^{-1}$. The behavior is similar
 to the one in \cite{RycerzPhD}.}
\label{fig:energy2H2}
\end{figure}

In Fig.~\ref{fig:energy2H2} we show the difference between the energies of the $\left( H_2 \right)_2$ system and that of the free molecules (per molecule):
\begin{align}
 \Delta E_{H_2} =  \frac{E_{\left( H_2 \right)_2}}{2}  - E_B,
 \label{eq:diffofenergy}
\end{align}
where $E_{\left( H_2 \right)_2}$ is the energy of the $\left( H_2 \right)_2$ system and $E_B=-2.29587 \ Ry$ is the energy of single molecule. The equilibrium parameters
are $\Delta E_{H_2} = -0.01129 \ Ry$ and $a = 4.5 \ a_0$. Those results are in agreement with the earlier estimates \cite{Kochanski}. The stability of hydrogen molecular clusters
were also studied in \cite{Hobza,Martinez}; our approach coherently incorporates electronic correlations (a necessity in describing the non-polar systems)
into the molecular picture, that plays an important role in view of the existence of the minimas of $\Delta E_{H_2} (a)$ curve \cite{Kochanski,Hobza}.

\section{Discussion of results and outlook}
\label{sec:results}

The evaluation of the global parameters of the system can be summarized as follows:
\begin{enumerate}
 \item	The $H_2$ binding energy is $E_B = -2.29587 \ Ry$. This value can be compared to the Ko\l{}os-Wolniewicz value \cite{Kolos}: $E_{\text{K-W}} = -2.349 \ Ry$,
	which is about $2.26 \%$ lower than the value obtained here.

 \item	The increase of the binding energy here is due to the zero-point motion, which is
 	\begin{equation}
 	  \label{eq:EZPM}
 	  E_{ZPM} = 0.024072 \ Ry,
 	\end{equation}
	and is of the order $1.0485 \%$ of the binding energy.

\item	Whereas the bond length is here $R_B = 1.43042 \ a_0$ (as compared to $R_{\text{K-W}} = 1.3984 \ a_0$, which is $ 2.29\%$ lower), the zero-point motion
	amplitude is $\left| \delta \vec{R} \right| = 0.189028 \ a_0$, a rather large value. Note that the optimal size of the inverse orbit of the $1s$
	component hydrogen orbital is $\alpha_B = 1.19378 \ a_0 ^{-1}$, so that the effective Bohr orbit is $a \equiv \alpha^{-1} = 0.8377 \ a_0$. The Bohr
	orbit decrease is due to the increased binding of electron in the molecule $\sim 0.2932 \ Ry$ with respect to that in the hydrogen atom. The size $a$
	is substantially smaller than that of $1s$ orbit ($1.06 \ \Ang$) in $H$ atom \cite{Spalek}.

\item	The ion--electron coupling constants versus $R$ are shown in Figs.~\ref{fig:couplings1} and \ref{fig:couplings2}, whereas their values for $R=R_B$ are listed in
	Table~\ref{tab:microscopic}. We also provide the second-order coupling constants values at the hydrogen-molecule equilibrium in the Table~\ref{tab:coupling}.
	Our first-principle calculations allow us to claim that the coupling constant appearing in the Holstein model \cite{Holstein}
	($\xi_\epsilon$) is decisively smaller than those of Su, Schrieffer and Heeger model \cite{Su} ($\xi_t$) as well as of those coming from the intersite
	direct Coulomb interaction ($\xi_K$). This should not be surprising, as the dominant coupling parameters represent interatomic-vibration contributions.
	A separate branch is represented by phonon excitations, but their analysis requires a construction of a spatially extended system of the molecules.
\end{enumerate}

The question is to what extent the calculated local characteristics will represent their local counterparts in the molecular-solid phase. Certainly, the phonons
and the molecule-rotational degrees of freedom will represent low-energy excitations. But the zero-point motion energy of the whole molecule should have to be added.
Will this provide a reliable description of the molecular or atomic hydrogen even though our accuracy in determining the individual-molecule energy is about
$2 \%$ higher than the virtually exact value of Ko\l{}os and Wolniewicz \cite{Kolos}? One has to check and this task is under consideration in our group.
Such consideration must include the inter-molecular hopping integrals $t_{13}$ and $t_{14}$.
One should also note that the proper method of treating the few-site $H_2$-molecule system is the quantum-Monte-Carlo-method \cite{Ceperley,Traynor,Azadi}.
Nonetheless, our method evaluates both the system energetics and the wave-function renormalization at the same time in the correlated state.

One can also extend the present model of the molecular binding by including also $2s$ and $2p$ adjustable hydrogen orbitals in the Hilbert space
of the single-particle states via the corresponding Gaussian representation. The first estimate of the 2s-orbital contribution to selected microscopic parameters is briefly discussed in~\ref{app:2sorbs}.
Their numerical  values are provided in ~\ref{tab:2s_equilibrium}. One can see that the basis extension leads to the numerically relevant corrections. This is an additional route to follow, but only after the first-principle calculations
of the solid phase along the lines discussed here is undertaken and tested.

Very recently \cite{Lee}, the dynamical mean field theory (DMFT) has been applied to $H_2$ molecule and its accuracy tested. Our approach in this respect is much simpler, but
still provides a comparable accuracy. Also, we have calculated the vibronic coupling  constants, which have been determined accurately recently \cite{Dickenson}. Those results
compare well with our estimates. This circumstance shows again that our method forms a proper starting point for treatment of solid molecular hydrogen, as a correlated
state, at least in the insulating phase.

\begin{table}
\caption{The numerical values of coefficients in Taylor series of ground-state energy. Up to the term $E_B^{(6)}$ all of the derivatives are calculated analytically
from equation \eqref{eq:Eg}. Orders seventh--ninth (marked by an asterisk) were calculated numerically
due to complicated analytical expression for ground-state energy.}
\label{tab:energies}
\begin{tabular}{||r|l||}\hline
$E_B^{(1)} \left( \frac{Ry}{a_0} \right)$ & \mc{1}{r||}{$0.0$}\\\hline
$\frac{1}{2!} E_B^{(2)} \left( \frac{Ry}{a_0 ^2} \right)$ & \mc{1}{r||}{$0.430045$}\\\hline
$\frac{1}{3!} E_B^{(3)} \left( \frac{Ry}{a_0 ^3} \right)$ & \mc{1}{r||}{$-0.464021$}\\\hline
$\frac{1}{4!} E_B^{(4)} \left( \frac{Ry}{a_0 ^4} \right)$ & \mc{1}{r||}{$0.354584$}\\\hline
$\frac{1}{5!} E_B^{(5)} \left( \frac{Ry}{a_0 ^5} \right)$ & \mc{1}{r||}{$-0.253393$}\\\hline
$\frac{1}{6!} E_B^{(6)} \left( \frac{Ry}{a_0 ^6} \right)$ & \mc{1}{r||}{$0.174863$}\\\hline\hline
$\frac{1}{7!} E_B^{(7)} \left( \frac{Ry}{a_0 ^7} \right) ^*$ & \mc{1}{r||}{$-0.119178$}\\\hline
$\frac{1}{8!} E_B^{(8)} \left( \frac{Ry}{a_0 ^8} \right) ^*$ & \mc{1}{r||}{$0.0817586$}\\\hline
$\frac{1}{9!} E_B^{(9)} \left( \frac{Ry}{a_0 ^9} \right) ^*$ & \mc{1}{r||}{$-0.0563837$}\\\hline
\end{tabular}
\end{table}

\begin{table}
  \caption{The values (in atomic units) of the microscopic parameters of Hamiltonian \eqref{eq:ham} and the electron--ion coupling constants
	  from \eqref{eq:coupling} at the hydrogen-molecule equilibrium ($R=R_B$ and $\alpha=\alpha_B$).}
 \label{tab:microscopic}
\begin{tabular}{||c|d||c|d||}\hline
 \mc{2}{||c||}{microscopic paramters ($Ry$)} & \mc{2}{c||}{coupling constants ($Ry/a_0$)} \\\hline\hline
 $\epsilon $ & -1.75079 & $\xi_\epsilon  $ & 0.00616165 \\\hline
 $t  $ & -0.727647 & $\xi_t $ & 0.598662 \\\hline
 $U  $ & 1.65321 & $\xi_U $ & -0.124934 \\\hline
 $K  $ & 0.956691 & $\xi_K $ & -0.234075 \\\hline
 $J  $ & 0.0219085 & $\xi_J $ & -0.00746303 \\\hline
 $V  $ & -0.0117991 & $\xi_V $ & -0.000426452 \\\hline
\end{tabular}
\end{table}

\begin{table}
  \caption{The values (in atomic units) of the second-order electron--ion coupling constants $\xi^2_i = {\delta ^2 \Xi}/{\delta R ^2}$ 
	  at the hydrogen-molecule equilibrium ($R=R_B$ and $\alpha=\alpha_B$).}
 \label{tab:coupling}
\begin{tabular}{||c|d||}\hline
  \mc{2}{||c||}{coupling constants ($Ry/a_0^2$)} \\\hline\hline
  $\xi^2_\epsilon  $ & 0.327335 \\\hline
  $\xi^2_t $ & -0.560426 \\\hline
  $\xi^2_U $ & 0.0504027 \\\hline
  $\xi^2_K $ & 0.013028 \\\hline
  $\xi^2_J $ & -0.00671566 \\\hline
  $\xi^2_V $ & -0.0105204 \\\hline
\end{tabular}
\end{table}

\section{Acknowledgments}
\label{sec:end}

The authors (APK \& JS) are grateful the Foundation for Polish Science (FNP) for financial support within the TEAM Project. M.A. is grateful to the Jagiellonian University
for hospitality and FNP for a partial support for his visit. J.S. is grateful to the National Science Center (NCN) for the support within the MAESTRO project, Grant
No.~DEC-2012/04/A/ST3/00342. APK visit at the University of Parma has been financially supported by the grant of the Polish Ministry of Science and Higher Education, Grant 
No.~7150/E-338/M/2013. M.M.M. acknowledges support by the Polish National Science Center (NCN) under grant No. DEC-2013/11/B/ST3/00824.

\appendix

\section{Exact solution without the zero-point motion}
\label{app:solution}

For the sake of completeness we express the microscopic parameters defined in \eqref{eq:microDefs} in terms of \textit{single-particle parameters} via \eqref{eq:wannier}
\begin{subequations}
 \begin{align}
\label{eq:eexpr}
\epsilon =& \beta ^2\left(1+\gamma ^2\right)\epsilon\Slater - 2\beta ^2\gamma \wyrozn{t}\Slater, 
\end{align}
\begin{align}
\label{eq:texpr}
\wyrozn{t} =& \beta ^2\left(1+\gamma ^2\right)\wyrozn{t}\Slater-2\beta ^2\gamma \epsilon\Slater, 
\end{align}
\begin{align}
\wyrozn{U} =& \beta ^4 \Big[ \left(1+\gamma ^4\right)\wyrozn{U}\Slater+2\gamma ^2 \wyrozn{K}\Slater \\\notag 
            &-4\gamma \left(1+\gamma ^2\right) \wyrozn{V}\Slater+4\gamma ^2 \wyrozn{J}\Slater \Big], 
\end{align}
\begin{align}
\wyrozn{K} =& \beta ^4  \Big[ 2\gamma ^2\wyrozn{U}\Slater+\left(1+\gamma ^4\right)\wyrozn{K}\Slater \\\notag
            &-4\gamma \left(1+\gamma ^2\right) \wyrozn{V}\Slater+4\gamma ^2 \wyrozn{J}\Slater \Big], 
\end{align}
\begin{align}
\wyrozn{J} =& \beta ^4  \Big[ 2\gamma ^2\wyrozn{U}\Slater+2\gamma ^2 \wyrozn{K}\Slater \\\notag
            &-4\gamma \left(1+\gamma ^2\right) \wyrozn{V}\Slater+\left(1+\gamma ^2\right)^2 \wyrozn{J}\Slater \Big], 
\end{align}
\begin{align}
\wyrozn{V} =& \beta ^4  \Big[ -\gamma \left(1+\gamma ^2\right)\wyrozn{U}\Slater-\gamma \left(1+\gamma ^2\right) \wyrozn{K}\Slater \\\notag
            &+\left(1+6\gamma ^2+\gamma ^4\right) \wyrozn{V}\Slater-2\gamma \left(1+\gamma ^2\right) \wyrozn{J}\Slater \Big],
 \end{align}
\end{subequations}
where $\Xi\Slater$ parameters are
\begin{subequations}
 \label{eq:microDefsApp}
 \begin{align}
  \label{eq:microDefsTApp}
    {T_{ij}}\Slater &= \matrixel{\Psi_i}{\mathcal{T}}{\Psi_j}, \\
  \label{eq:microDefsVApp}
    {V_{ijkl}}\Slater &= \matrixel{\Psi_i \Psi_j}{\mathcal{V}_{12}}{\Psi_k \Psi_l},
 \end{align}
\end{subequations}
with $\mathcal{T} = - \bigtriangledown^2 - {2} / | \vec{r} - \vec{R} |$, and $\mathcal{V} =  {2} / | \vec{r} - \vec{r}' |$. \textit{The single-particle parameters} read
\begin{subequations}
\label{eq:SPP}
 \begin{align}
  \wyrozn{\epsilon}\Slater =&\alpha ^2-2\alpha -\frac{2}{R}+2\left(\alpha +\frac{1}{R}\right) e^{-2\alpha  R}, 
\end{align}
\begin{align}
  \wyrozn{t}\Slater =&\alpha ^2 e^{-\alpha  R} \left(1+\alpha  R - \frac{1}{3}\alpha ^2R^2\right) \\\notag
       &-4\alpha  e^{-\alpha  R}(1+ \alpha  R),
\end{align}
\begin{align}
  \wyrozn{U}\Slater =&\frac{5}{4}\alpha,
\end{align}
\begin{align}
  \wyrozn{K}\Slater =&\frac{2}{R}-\alpha e^{-2\alpha  R} \left(\frac{2}{\alpha  R}+\frac{3}{2}\alpha  R+\frac{1}{3}\alpha ^2R^2+\frac{11}{4}\right),
\end{align}
\begin{align}
  \wyrozn{V}\Slater =&\alpha e^{-\alpha  R}\left(2 \alpha  R+\frac{1}{4}+\frac{5}{8\alpha  R}\right) \\\notag
	&-\frac{1}{4} \alpha e^{-3 \alpha  R} \left(1+\frac{5}{2 \alpha  R}\right), 
\end{align}
\begin{align}
  \wyrozn{J}\Slater =&\alpha  e^{-2 \alpha  R}\left(\frac{5}{4}-\frac{23}{10}\alpha  R-\frac{6}{5}\alpha ^2R^2-\frac{2}{15}\alpha ^3R^3\right)\\\notag
	&+\frac{12}{5R} \bigg( S^2 C_E +S^2 \log(\alpha  R) - 2 S \bar{S} \text{Ei}(-2 \alpha  R) \\\notag
        &+\bar{S}^2 \text{Ei}(-4 \alpha  R) \bigg),
 \end{align}
\end{subequations}
where the overlaps are given by
\begin{align}
 \label{eq:overlap}
  S = e^{-\alpha  R}\left(1+\alpha  R + \frac{1}{3}\alpha ^2 R^2\right), \\
 \bar{S} = e^{\alpha  R}\left(1-\alpha  R + \frac{1}{3}\alpha ^2 R^2\right).
\end{align}
$C_E$ is so-called Euler constant
\begin{align}
\label{eq:defs}
 C_E = \lim_{n \rightarrow \infty } \left(\sum_{k=1}^n \frac{1}{k} - \log(n) \right) \approx 0.5772,\\ 
\end{align}
and $Ei(x)$ is the Exponential Integral.
\begin{align}
 \label{eq:expInt}
 Ei(x) = - \int_{-x}^{\infty} e^{-t}t^{-1} dt.
\end{align}

\section{Adiabatic-approximation details}
\label{app:adiabatic}

For the sake of completeness, we provide also the explicit form of the coupling constants $\xi_i \equiv d \Xi_i/d R$. For editorial purposes we 
calculate first \textit{the single-particle coupling constants} from \eqref{eq:SPP}. They are
\begin{subequations}
 \begin{align}
  \wyrozn{\delta \epsilon}\Slater =& 2 R^{-2} e^{-2 \alpha R} \left[e^{2 \alpha R} - 1 -2 R \alpha  (1+ \alpha R)\right],
\end{align}
\begin{align}
  \wyrozn{\delta t}\Slater =& \frac{1}{3} e^{- \alpha R} \alpha ^3 R [12+\alpha  (-5+ \alpha R)], \\
  \wyrozn{\delta U}\Slater =& 0.
 \end{align}
The corresponding derivatives of the two-particle parameters are
\begin{widetext}
 \begin{align}
  \wyrozn{\delta K}\Slater =& \frac{1}{3} R^{-2} e^{-2 \alpha R } \Big[ 6-6 e^{2 \alpha R} + \alpha R \big( 2+ \alpha R \big) \big(6+ \alpha R (3+2 \alpha R) \big) \Big] \\
  \wyrozn{\delta V}\Slater =& \frac{1}{8} R^{-2} \bigg( e^{-3 R \alpha } \Big[ 5 + 15 R \alpha + 6 R^2 \alpha^2  \Big] -e^{- R \alpha } \Big[ 5+5 R \alpha -14 R^2 \alpha ^2+16 R^3 \alpha ^3 \Big] \bigg) \\
 \wyrozn{\delta J}\Slater =& \frac{1}{15 R^2}e^{-2 R \alpha } \bigg(-4 C_E \Big[ 9+18 R \alpha +21 R^2 \alpha ^2+18 R^3 \alpha ^3+9 R^4 \alpha ^4+2 R^5 \alpha ^5 \Big] \\\notag
&+R^2 \alpha ^2 \Big[72+33 R \alpha +30 R^2 \alpha ^2+4 R^3 \alpha ^3\Big] +4 e^{4 R \alpha } \Big[ -9+18 R \alpha -21 R^2 \alpha ^2+18 R^3 \alpha ^3-9 R^4 \alpha ^4+2 R^5 \alpha ^5 \Big] \text{Ei}(-4 R \alpha) \\\notag
&-24 e^{2 R \alpha } \left(-3-R^2 \alpha ^2+R^4 \alpha ^4\right) \text{Ei}(-2 R \alpha) -36 \log(R \alpha ) -4 R \alpha  (18+21 R \alpha +18 R^2 \alpha ^2+9 R^3 \alpha ^3+2 R^4 \alpha ^4) \log (R \alpha ) \bigg)
 \end{align}
\end{widetext}
\end{subequations}
with basis mixing-parameters $\beta$ and $\gamma$ (cf. \eqref{eq:mix}) changes
\begin{align}
 \delta \beta = \frac{S}{4(1-S^2) \beta} \left( \frac{1}{\sqrt{1-S^2}} + \frac{1}{2(1-S^2)}\right) \delta S, \\
 \delta \gamma = \frac{1}{1-S^2+\sqrt{1-S^2}} \delta S,
\end{align}
where $S$ is the overlap \eqref{eq:overlap} and $\delta S$ reads
\begin{align}
 \delta S = -\frac{1}{3} e^{-R \alpha } R \alpha ^2 (1+R \alpha ).
\end{align}

Our final results are
\begin{subequations}
 \begin{align}
\xi_\epsilon = 2 \frac{\delta \beta}{\beta} \epsilon &+ \beta ^2 \Big[ ( 1+\gamma ^2 ) \delta \epsilon\Slater - 2 \gamma \wyrozn{\delta t}\Slater \Big] \\\notag
&+ 2\beta ^2 \delta \gamma \Big[ \gamma \epsilon\Slater -  \wyrozn{t}\Slater \Big],
\end{align}
\begin{align}
\wyrozn{\xi_ t} = 2 \frac{\delta \beta}{\beta} \wyrozn{t} &+ \beta ^2 \Big[ ( 1+\gamma ^2 ) \wyrozn{\delta t}\Slater - 2 \gamma \delta \epsilon\Slater \Big] \\\notag
&+ 2\beta ^2 \delta \gamma \Big[ \gamma \wyrozn{t}\Slater -  \epsilon\Slater \Big],
\end{align}
\begin{align}
\wyrozn{\xi_ U} = 4 \frac{\delta \beta}{\beta} \wyrozn{U} &+ \beta ^4 \Big[ \left(1+\gamma ^4\right)\wyrozn{\delta U}\Slater+2\gamma ^2 \wyrozn{\delta K}\Slater  \\\notag
&-4\gamma \left(1+\gamma ^2\right) \wyrozn{\delta V}\Slater+4\gamma ^2 \wyrozn{\delta J}\Slater \Big] \\\notag
&+ 4\beta ^4 \delta \gamma \Big[\gamma ^3 \wyrozn{U}\Slater +  \gamma \wyrozn{K}\Slater  \\\notag
&-( 1+3 \gamma^2 ) \wyrozn{V}\Slater + 2 \gamma \wyrozn{J}\Slater \Big],
\end{align}
\begin{align}
\wyrozn{\xi_ K} = 4 \frac{\delta \beta}{\beta} \wyrozn{K} &+ \beta ^4\Big[2\gamma ^2\wyrozn{\delta U}\Slater+\left(1+\gamma ^4\right)\wyrozn{\delta K}\Slater  \\\notag
&-4\gamma \left(1+\gamma ^2\right) \wyrozn{\delta V}\Slater+4\gamma ^2 \wyrozn{\delta J}\Slater\Big] \\\notag
&+ 4\beta ^4 \delta \gamma \Big[ \gamma\wyrozn{U}\Slater + \gamma ^3 \wyrozn{K}\Slater \\\notag
&-( 1+3 \gamma^2 ) \wyrozn{V}\Slater + 2 \gamma \wyrozn{J}\Slater \Big],
\end{align}
\begin{align}
\wyrozn{\xi_ J} = 4 \frac{\delta \beta}{\beta} \wyrozn{J} &+ \beta ^4\Big[2\gamma ^2\wyrozn{\delta U}\Slater+2\gamma ^2 \wyrozn{\delta K}\Slater \\\notag
&-4\gamma \left(1+\gamma ^2\right) \wyrozn{\delta V}\Slater+\left(1+\gamma ^2\right)^2 \wyrozn{\delta J}\Slater\Big] \\\notag
&+ 4\beta ^4 \delta \gamma \Big[ \gamma \wyrozn{U}\Slater +  \gamma \wyrozn{K}\Slater \\\notag
&-( 1+3 \gamma^2 ) \wyrozn{V}\Slater +  \gamma (1 + \gamma ^2)\wyrozn{J}\Slater \Big],
\end{align}
\begin{align}
\wyrozn{\xi_ V} = 4 \frac{\delta \beta}{\beta} \wyrozn{V} &- \beta ^4\Big[\gamma \left(1+\gamma ^2\right)\wyrozn{\delta U}\Slater \\\notag
&+\gamma \left(1+\gamma ^2\right) \wyrozn{\delta K}\Slater -\left(1+6\gamma ^2+\gamma ^4\right) \wyrozn{\delta V}\Slater\\\notag
&+2\gamma \left(1+\gamma ^2\right) \wyrozn{\delta J}\Slater\Big] \\\notag
&- \beta ^4 \delta \gamma \Big[(1+3 \gamma ^2)\wyrozn{U}\Slater +(1+3 \gamma ^2)\wyrozn{K}\Slater  \\\notag
&- 4 \gamma (3 + \gamma ^2) \wyrozn{V}\Slater +2 (1+3 \gamma ^2)\wyrozn{J}\Slater \Big].
 \end{align}
\end{subequations}

These parameters are displayed vs. $R$ in~Figures \ref{fig:couplings1} and \ref{fig:couplings2}.

\section{Zero-point motion with classical electronic interaction}

We ask the question how important is to include the quantum nature of the electronic interaction in our evaluation of zero-point motion energy.
Let us consider, following \cite{Spalek}, the energy of the ions as
\begin{align}
 \label{eq:EnIon}
 E_{ion} = \frac{\delta P ^2}{2 M} + \frac{e^2}{R + \delta R},
\end{align}
where ${\delta P}$ and ${\delta R}$ are the momentum and position uncertainties, $M$ is ion mass and $e$ its charge. By expressing ${\delta P}$ by ${\delta R}$ via
uncertainty relation ${\delta P}^2{\delta R}^2 \geqslant \frac{3}{4} \hbar^2$ we obtain
\begin{align}
 \label{eq:EnIon}
 E_{ion} = \frac{\frac{3}{4} \hbar^2}{2 M \delta R ^2} + \frac{e^2}{R + \delta R},
\end{align}
that we can minimize with respect to $\delta R$. We calculate

\begin{align}
\delta R _0  \stackrel{a.u.}{=} \frac{1}{4 \sqrt{2} M}\left(a - 1 + \frac{1-8 \sqrt{2} M R}{a} \right),
\end{align}
where
\begin{align}
 a \stackrel{a.u.}{=}& \Big[-1+12 M R \left(\sqrt{2}-4 M R\right) + \\\notag
      &2^{15/4} \sqrt{M^3 R^3 \left(-1+9 \sqrt{2} M R\right)} \Big]^{1/3}.
\end{align}

\begin{table}
  \caption{The values (in atomic units) of the zero-point motion energy and amplitude. The classical electron interaction approximation versus adiabatic approximation.}
 \label{tab:zpm}
\begin{tabular}{||r||d|d||}\hline
\mc{1}{||c||}{$ $} & \mc{1}{c|}{$|\delta \vec{R}_0|$ ($a_0$)} & \mc{1}{c||}{$E_{ZPM}$ ($Ry$) } \\\hline\hline
classical interaction & 0.0901816 & 0.14434 \\\hline
quantum interaction & 0.189028 & 0.024072 \\\hline
\end{tabular}
\end{table}

We take the mass of the ion $M \stackrel{a.u.}{\approx} 1836.15267 m_e$ and the interionic distance $R=R_B = 1.43017 a_0$. The results are presented in the Table~\ref{tab:zpm}.

\section{Inclusion of $2s$ orbitals}
\label{app:2sorbs}

We would like to estimate the role of higher orbitals both for more realistic description of $H_2$ systems and future consideration of other elements.
We start by taking the $2s$ Slater-type orbital
\begin{align}
  \label{eq:slater2s}
 \Psi ^{2s} \left(\vec{r} \right) \equiv \sqrt{\frac{\alpha_{2s}^5}{3 \pi }} | \vec{r} | e^{- \alpha_{2s}  | \vec{r} | },
\end{align}
where $\alpha_{2s}$ is the inverse wave function size. It is obviously non-orthogonal with the $1s$ orbital \eqref{eq:slater} as we have that
\begin{align}
\label{eq_S_on}
 S_{on}^{\alpha, \alpha_{2s}} \equiv \bracket{\Psi ^{1s}}{\Psi ^{2s}} = \frac{8 \sqrt{3} \alpha ^{3/2} \alpha_{2s} ^{5/2}}{(\alpha +\alpha_{2s} )^4}.
\end{align}

\subsection{On-site orthogonalization}
We perform the orthogonalization by introducing realistic orbital functions \cite{Calderini}
\begin{align}
 \label{eq:O-STO}
 \chi ^{1s} \left( \vec{r} \right) &=   \Psi ^{1s} \left(\alpha, \vec{r} \right), \\
 \chi ^{2s} \left( \vec{r} \right) &= A \Psi ^{1s} \left(\alpha_{2s}, \vec{r} \right) + B \Psi ^{2s} \left(\alpha_{2s}, \vec{r} \right),
\end{align}
where $A$ and $B$ are mixing parameters obtained via orthonormality conditions
\begin{align}
 \label{eq:o_con_1}
 \bracket{\chi ^{1s}}{\chi ^{2s}} = 0, \\\notag
 \bracket{\chi ^{2s}}{\chi ^{2s}} = 1.
\end{align}

We can solve problem \eqref{eq:o_con_1} analytically and obtain
\begin{align}
 \label{eq_o_2s_solve}
 A &= -\frac{{ S_{on}^{\alpha, \alpha_{2s}}}}{\sqrt{{ S_{1s}^{\alpha, \alpha_{2s}}}^2-2 { S_{1s}^{\alpha, \alpha_{2s}}} { S_{on}^{\alpha, \alpha_{2s}}} {S_{on}^{\alpha_{2s}, \alpha_{2s}}}+{ S_{on}^{\alpha, \alpha_{2s}}}^2}}, \\\notag
 B &=  \frac{{ S_{1s}^{\alpha, \alpha_{2s}}}}{\sqrt{{ S_{1s}^{\alpha, \alpha_{2s}}}^2-2 { S_{1s}^{\alpha, \alpha_{2s}}} { S_{on}^{\alpha, \alpha_{2s}}} {S_{on}^{\alpha_{2s}, \alpha_{2s}}}+{ S_{on}^{\alpha, \alpha_{2s}}}^2}},
\end{align}
where $S_{on}$ is given by \eqref{eq_o_2s_solve} and
\begin{align}
\label{eq_S_on}
 S_{1s}^{\alpha, \alpha_{2s}} \equiv \bracket{\Psi ^{1s}}{\Psi ^{1s}} =\frac{8 (\alpha  \alpha_{2s} )^{3/2}}{(\alpha +\alpha_{2s} )^3}.
\end{align}

\subsection{Intersite Orthogonalization}
As $\chi^\sigma$'s are orthogonal on-site, one can also introduce intersite orthogonalization.
We introduce the following mixing coefficients \eqref{eq:wannier}
\begin{align}
 \label{eq:wannier_sigma}
 w^\sigma _i \left(  \vec{r} \right) = \beta^\sigma ( \chi^\sigma_i \left(  \vec{r} \right) - \gamma^\sigma \chi_j \left(  \vec{r} \right) ), 
\end{align}
where $\beta^\sigma$ and $\gamma^\sigma$ depend only on the overlap integral $S^\sigma \equiv \bracket{\chi^\sigma _1}{\chi^\sigma _2}$:
\begin{align}
\label{eq:mix_sigma}
  \beta^\sigma = \frac{1}{\sqrt{2}}\sqrt{\frac{1+\sqrt{1-(S^\sigma)^2}}{1-(S^\sigma)^2}} \\
  \gamma^\sigma = \frac{S^\sigma}{1+\sqrt{1-(S^\sigma)^2}}.
 \end{align}
 
 We already have overlap $S^{1s}$ \eqref{eq:overlap}
 
 Overlap $S^{2s}$ is only a little bit more complicated
 \begin{align}
  S^{2s} &= \bracket{\chi^{2s} _1}{\chi^{2s} _2} =  \bracket{A \Psi ^{1s} _1 + B \Psi ^{2s} _1}{A \Psi ^{1s} _2 + B \Psi ^{2s} _2} = \\\notag
         &= A^2 \bracket{\Psi^{1s} _1}{\Psi^{1s} _2} + 2 A B \bracket{\Psi^{1s} _1}{\Psi^{2s} _2} + B^2 \bracket{\Psi^{2s} _1}{\Psi^{2s} _2} = \\\notag
         &= A^2 S^{1s,\alpha_{2s}} + 2 A B S^{1s,2s} + B^2 S^{2s},
 \end{align}
 where
 \begin{align}
  S^{1s,2s} &= \frac{e^{{\alpha_{2s}}  (-R)} ({\alpha_{2s}}  R ({\alpha_{2s}}  R ({\alpha_{2s}}  R+4)+9)+9)}{6 \sqrt{3}}, \\
  S^{2s}    &= \frac{1}{45} e^{\alpha_{2s}  (-R)} (\alpha_{2s}  R (\alpha_{2s}  R (\alpha_{2s}  R (\alpha_{2s}  R\\\notag
	      &+5)+20)+45)+45).
 \end{align}
 
 \subsection{Single-particle microscopic parameters}
 
Introducing $2s$ orbitals provides us with four new single-particle microscopic parameters
\begin{subequations}
 \label{eq:2sMP_defs}
\begin{align}
 \epsilon_{2s} &= \matrixel{w ^{2s} _i \left( \vec{r} \right) }{\mathcal{H}_1}{w ^{2s} _i \left( \vec{r} \right) }, \\
 t_{2s}        &= \matrixel{w ^{2s} _i \left( \vec{r} \right) }{\mathcal{H}_1}{w ^{2s} _{\bar{i}} \left( \vec{r} \right) }, \\
 V_{on}        &= \matrixel{w ^{1s} _i \left( \vec{r} \right) }{\mathcal{H}_1}{w ^{2s} _i \left( \vec{r} \right) }, \\
 V_{inter}     &= \matrixel{w ^{1s} _i \left( \vec{r} \right) }{\mathcal{H}_1}{w ^{2s} _{\bar{i}} \left( \vec{r} \right) },
\end{align}
\end{subequations}
 where $\epsilon_{2s}$ is single-particle energy on $2s$ orbital, $t_{2s}$ the hopping between $2s$ sites, and $V_{on}$ and $V_{inter}$ are
 hybridizations on-- and inter-site respectively.
 
 Similarly to Sec. \ref{app:solution} and \ref{app:adiabatic}, the exact solution is a function of Slater microscopic parameters
 \begin{widetext}
 \begin{subequations}
  \label{eq:2sMP_funs}
  \begin{align}
   \epsilon_{2s} &= {\beta^{2s}} ^2 \left(A^2 \left(\left(\gamma ^2+1\right) {\epsilon_{1s} '}-2 \gamma {t_{1s} '}\right)+2 A B \left(\left(\gamma ^2+1\right) {V_{on} '}-2 \gamma  {V_{inter} '}\right)+B^2 \left(\left(\gamma ^2+1\right) {\epsilon_{2s} '}-2 \gamma  {t_{2s} '}\right)\right), \\
   t_{2s}        &= {\beta^{2s}} ^2 \left(A^2 \left(\left(\gamma ^2+1\right){t_{1s} '}-2 \gamma  {\epsilon_{1s} '}\right)+2 A B \left(\left(\gamma ^2+1\right) {V_{inter} '}-2 \gamma  {V_{on} '}\right)+B^2 \left(\left(\gamma ^2+1\right) {t_{2s} '}-2 \gamma  {\epsilon_{2s} '}\right)\right), \\
   V_{on}        &= {\beta^{1s}} {\beta^{2s}} \left(A \left(\left({ \gamma^{1s}}  { \gamma^{2s}} +1\right) {\epsilon_{1s} ''}-({ \gamma^{1s}} + { \gamma^{2s}} )  {t_{1s} ''}\right)+B \left(\left({ \gamma^{1s}}  { \gamma^{2s}} +1\right) {V_{on} ''}-({ \gamma^{1s}} + { \gamma^{2s}} )  {V_{inter} ''}\right)\right), \\
   V_{inter}     &= {\beta^{1s}} {\beta^{2s}} \left(A \left(\left({ \gamma^{1s}}  { \gamma^{2s}} +1\right) {t_{1s} ''}-({ \gamma^{1s}} + { \gamma^{2s}} )  {\epsilon_{1s} ''}\right)+B \left(\left({ \gamma^{1s}}  { \gamma^{2s}} +1\right) {V_{inter} ''}-({ \gamma^{1s}} + { \gamma^{2s}} )  {V_{on} ''}\right)\right),
  \end{align}
 \end{subequations}
 \end{widetext}
 where $A$ and $B$ are found via \eqref{eq:o_con_1}, while $\beta^{\sigma}$ and $\gamma^{\sigma}$ via \eqref{eq:mix_sigma}. The Slater microscopic parameters
 can be explicitly written in a form
 \begin{widetext}
 \begin{align}
  {\epsilon_{1s} '} &= \frac{1}{R} {e^{-2 { \alpha_{2s} }  R} (2 { \alpha_{2s} }  R+e^{2 { \alpha_{2s} }  R} (({ \alpha_{2s} } -2) { \alpha_{2s} }  R-2)+2)} 
 \end{align}
 \begin{align}
  {t_{1s} '} &= -\frac{1}{3} { \alpha_{2s} }  e^{{ \alpha_{2s} }  (-R)} ({ \alpha_{2s} }  (R ({ \alpha_{2s} }  ({ \alpha_{2s} }  R-3)+12)-3)+12),
 \end{align}
 \begin{align}
  {\epsilon_{2s} '} &= \frac{1}{3 R} e^{-2 { \alpha_{2s} }  R} (e^{2 { \alpha_{2s} }  R} (({ \alpha_{2s} } -3) { \alpha_{2s} }  R-6)+{ \alpha_{2s} }  R (2 { \alpha_{2s} }  R ({ \alpha_{2s} }  R+3)+9)+6),
 \end{align}
 \begin{align}
  {t_{2s} '} &= -\frac{1}{45} { \alpha_{2s} }  e^{{ \alpha_{2s} }  (-R)} ({ \alpha_{2s} }  (R ({ \alpha_{2s} }  (R ({ \alpha_{2s} }  R ({ \alpha_{2s} }  ({ \alpha_{2s} }  R-5)+10)+40)-15)+90)-15)+90),
 \end{align}
 \begin{align}
  {V_{on} '} &= \frac{1}{2 \sqrt{3} R} e^{-2 { \alpha_{2s} }  R} (4 { \alpha_{2s} } ^2 R^2+8 { \alpha_{2s} }  R+e^{2 { \alpha_{2s} }  R} (({ \alpha_{2s} } -4) { \alpha_{2s} }  R-6)+6),
 \end{align}
 \begin{align}
  {V_{inter} '} &= -\frac{1}{6 \sqrt{3}} { \alpha_{2s} }  e^{{ \alpha_{2s} }  (-R)} ({ \alpha_{2s} } ^4 R^3-4 { \alpha_{2s} } ^3 R^2+3 { \alpha_{2s} } ^2 R (4 R-1)+3 { \alpha_{2s} }  (8 R-1)+24),
 \end{align}
 \begin{align}
  {\epsilon_{1s} ''} &= \frac{8 (\alpha  { \alpha_{2s} } )^{3/2}}{R (\alpha +{ \alpha_{2s} } )^3} e^{-R (\alpha +{ \alpha_{2s} } )} (R (\alpha +{ \alpha_{2s} } )+e^{R (\alpha +{ \alpha_{2s} } )} ((\alpha -1) { \alpha_{2s} }  R+\alpha  (-R)-2)+2),
 \end{align}
 \begin{align}
  {t_{2s} ''} &= -\frac{8 (\alpha  { \alpha_{2s} } )^{3/2}}{R (\alpha ^2-{ \alpha_{2s} } ^2)^3} e^{-R (\alpha +{ \alpha_{2s} } )} (e^{\alpha  R} (-2 \alpha ^2 (\alpha +1) { \alpha_{2s} } -2 (\alpha -1) { \alpha_{2s} } ^3+\alpha ^3 (\alpha  R+2) \\\notag
	      &+(\alpha -1) { \alpha_{2s} } ^4 (-R)+\alpha  { \alpha_{2s} } ^2 ((\alpha -2) \alpha  R-2))+e^{{ \alpha_{2s} }  R} (2 \alpha ^3 ({ \alpha_{2s} } -1)+2 \alpha  { \alpha_{2s} } ^2 ({ \alpha_{2s} } +1) \\\notag
	      &+\alpha ^4 ({ \alpha_{2s} } -1) R+\alpha ^2 { \alpha_{2s} }  (2-({ \alpha_{2s} } -2) { \alpha_{2s} }  R)-{ \alpha_{2s} } ^3 ({ \alpha_{2s} }  R+2))),
 \end{align}
 \begin{align}
  {V_{on} ''}&= \frac{8 \alpha ^{3/2} { \alpha_{2s} } ^{5/2}}{\sqrt{3} R (\alpha +{ \alpha_{2s} } )^4} e^{-R (\alpha +{ \alpha_{2s} } )} (R (\alpha +{ \alpha_{2s} } ) (R (\alpha +{ \alpha_{2s} } )+4) \\\notag
	     &-e^{R (\alpha +{ \alpha_{2s} } )} (R (\alpha  (\alpha -2 { \alpha_{2s} } +2)+2 { \alpha_{2s} } )+6)+6),
 \end{align}
 \begin{align}
  {V_{inter} ''}&= \frac{8 \alpha ^{3/2} { \alpha_{2s} } ^{5/2}}{\sqrt{3} R (\alpha -{ \alpha_{2s} } )^4 (\alpha +{ \alpha_{2s} } )^4} e^{-5 \alpha  R-4 { \alpha_{2s} }  R} (e^{4 R (\alpha +{ \alpha_{2s} } )} (2 \alpha ^5+8 \alpha ^3 { \alpha_{2s} }  (2 { \alpha_{2s} } -1)+2 \alpha  { \alpha_{2s} } ^3 (3 { \alpha_{2s} } +4) \\\notag
		&+\alpha ^6 R +2 \alpha ^4 (({ \alpha_{2s} } -1) { \alpha_{2s} }  R+1)+\alpha ^2 { \alpha_{2s} } ^2 ({ \alpha_{2s} }  (4-3 { \alpha_{2s} } ) R+4)-2 { \alpha_{2s} } ^4 ({ \alpha_{2s} }  R+3)) \\\notag
		&-e^{5 \alpha  R+3 { \alpha_{2s} }  R} ((\alpha -1) { \alpha_{2s} } ^6 R^2+\alpha ^4 (\alpha  (R (\alpha  R+2)+2)+2)-4 \alpha ^3 { \alpha_{2s} }  (\alpha  (\alpha +1) R+2) \\\notag
		&+\alpha ^2 { \alpha_{2s} } ^2 (\alpha  (R ((\alpha -3) \alpha  R-4)+16)+4)+4 (\alpha -1) { \alpha_{2s} } ^5 R+{ \alpha_{2s} } ^4 (\alpha  ( R (\alpha  (3-2 \alpha ) R+2)+6)-6)+8 \alpha  { \alpha_{2s} } ^3 (\alpha  R+1))).
  \end{align}
 \end{widetext}

 One can obtain the exact values for the optimal inter-ionic distance $R=R_B=1.43042$ and $\alpha =\alpha_B = 1.19378$. The results, together  with comparison
 to the one-orbital case, are listed in Table~\ref{tab:2s_equilibrium}. Note that the new estimates are carried out for the optimal bond length and the inverse wave-functions size 
 for the case of 1s functions only  ($R=R_B=1.43042$ and $\alpha =\alpha_B = 1.19378$, respectively).
 
 \begingroup
\squeezetable
\begin{table*}[h]
 \caption{\label{tab:2s_equilibrium} The values of single-particle microscopic parameters without optimization of $2s$ orbitals
 for $1s$ and $2s$ band ($R=R_B$, $\alpha =\alpha_B$, $\alpha_{2s} =\alpha_B/2$). Note that last column describes
 atomic limit, where $\alpha \rightarrow 1$, $\alpha_{2s} \rightarrow 0.5$. We can observe that the model fulfills requirements, as the Wannier function
 $w^{2s}_i$ approaches the exact solution of Hydrogen atom.}
\begin{tabular}{|d||d|d|d||d|}
\cline{2-5}
\mc{1}{c||}{$ $} & \mc{3}{c||}{Equilibrium system} & \mc{1}{c|}{Atomic limit} \\\hline
\mc{1}{|c||}{$ $} &\mc{1}{c|}{$ $} &\mc{1}{c|}{$ $} &\mc{1}{c||}{$ $} &\mc{1}{c|}{$ $}  \\
\mc{1}{|c||}{microscopic} &\mc{1}{c|}{$\Xi^{2s} (Ry) $} &\mc{1}{c|}{$\Xi^{1s} (Ry) $} &\mc{1}{c||}{$\Xi^{2s}$/$\Xi^{1s}$} &\mc{1}{c|}{$\Xi^{2s} (Ry) $}  \\
\mc{1}{|c||}{parameter $\Xi$} &\mc{1}{c|}{$ $} &\mc{1}{c|}{$ $} &\mc{1}{c||}{$ $} &\mc{1}{c|}{$ $}  \\\hline\hline
\mc{1}{|c||}{$\epsilon$} & -0.518585 & -1.75079 & 29.62 \% & -0.25 \\\hline
\mc{1}{|c||}{$t$} & -0.292465 & -0.727647 & 40.19 \% & 0 \\\hline
\mc{1}{|c||}{$V_{on}$} & 0.0773174 & \mc{2}{c||}{$ $} & 0 \\\cline{1-2}\cline{5-5}
\mc{1}{|c||}{$V_{inter}$} & -0.110457 & \mc{2}{c||}{$ $} & 0 \\\cline{1-2}\cline{5-5}
\end{tabular}
\end{table*}
\endgroup

\bibliography{bibliography}

\end{document}